# *Euclid* preparation

## LX. The use of HST images as input for weak-lensing image simulations

Euclid Collaboration: D. Scognamiglio[⋆,1,2], T. Schrabback[3,2], M. Tewes[2], B. Gillis[4], H. Hoekstra[5], E. M. Huff[1], O. Marggraf[2], T. Kitching[6], R. Massey[7,8], I. Tereno[9,10], C. S. Carvalho[10], A. Robertson[1], G. Congedo[4], N. Aghanim[11], B. Altieri[12], A. Amara[13], S. Andreon[14], N. Auricchio[15], C. Baccigalupi[16,17,18,19], M. Baldi[20,15,21], S. Bardelli[15], P. Battaglia[15], C. Bodendorf[22], D. Bonino[23], E. Branchini[24,25,14], M. Brescia[26,27,28], J. Brinchmann[29,30], S. Camera[31,32,23], V. Capobianco[23], C. Carbone[33], V. F. Cardone[34,35], J. Carretero[36,37], S. Casas[38], F. J. Castander[39,40], M. Castellano[34], G. Castignani[15], S. Cavuoti[27,28], A. Cimatti[41], C. Colodro-Conde[42], C. J. Conselice[43], L. Conversi[44,12], Y. Copin[45], F. Courbin[46], H. M. Courtois[47], M. Cropper[6], A. Da Silva[9,48], H. Degaudenzi[49], G. De Lucia[17], A. M. Di Giorgio[50], J. Dinis[9,48], F. Dubath[49], C. A. J. Duncan[43], X. Dupac[12], S. Dusini[51], M. Farina[50], S. Farrens[52], S. Ferriol[45], P. Fosalba[40,53], M. Frailis[17], E. Franceschi[15], S. Galeotta[17], C. Giocoli[15,54], P. Gómez-Alvarez[55,12], A. Grazian[56], F. Grupp[22,57], L. Guzzo[58,14], S. V. H. Haugan[59], W. Holmes[1], F. Hormuth[60], A. Hornstrup[61,62], P. Hudelot[63], K. Jahnke[64], B. Joachimi[65], E. Keihänen[66], S. Kermiche[67], A. Kiessling[1], M. Kilbinger[52], B. Kubik[45], M. Kümmel[57], M. Kunz[68], H. Kurki-Suonio[69,70], S. Ligori[23], P. B. Lilje[59], V. Lindholm[69,70], I. Lloro[71], G. Mainetti[72], E. Maiorano[15], O. Mansutti[17], K. Markovic[1], M. Martinelli[34,35], N. Martinet[73], F. Marulli[74,15,21], E. Medinaceli[15], S. Mei[75], Y. Mellier[76,63], M. Meneghetti[15,21], G. Meylan[46], M. Moresco[74,15], L. Moscardini[74,15,21], R. Nakajima[2], S.-M. Niemi[77], J. W. Nightingale[8], C. Padilla[78], S. Paltani[49], F. Pasian[17], K. Pedersen[79], S. Pires[52], G. Polenta[80], M. Poncet[81], L. A. Popa[82], L. Pozzetti[15], F. Raison[22], R. Rebolo[42,83], A. Renzi[84,51], J. Rhodes[1], G. Riccio[27], E. Romelli[17], M. Roncarelli[15], E. Rossetti[20], R. Saglia[57,22], Z. Sakr[85,86,87], A. G. Sánchez[22], D. Sapone[88], B. Sartoris[57,17], R. Scaramella[34,35], M. Schirmer[64], P. Schneider[2], A. Secroun[67], G. Seidel[64], S. Serrano[40,89,39], C. Sirignano[84,51], G. Sirri[21], J. Skottfelt[90], L. Stanco[51], J.-L. Starck[52], J. Steinwagner[22], P. Tallada-Crespí[36,37], A. N. Taylor[4], H. I. Teplitz[91], R. Toledo-Moreo[92], F. Torradeflot[37,36], I. Tutusaus[86], L. Valenziano[15,93], T. Vassallo[57,17], G. Verdoes Kleijn[94], A. Veropalumbo[14,25,95], Y. Wang[91], J. Weller[57,22], M. Wetzstein[22], G. Zamorani[15], E. Zucca[15], A. Biviano[17,16], M. Bolzonella[15], A. Boucaud[75], E. Bozzo[49], C. Burigana[96,93], M. Calabrese[97,33], J. A. Escartin Vigo[22], J. Gracia-Carpio[22], N. Mauri[41,21], A. Pezzotta[22], M. Pöntinen[69], C. Porciani[2], V. Scottez[76,98], M. Tenti[21], M. Viel[16,17,19,18,99], M. Wiesmann[59], Y. Akrami[100,101], V. Allevato[27], S. Anselmi[51,84,102], M. Ballardini[103,15,104], L. Blot[105,102], S. Borgani[106,16,17,18], S. Bruton[107], R. Cabanac[86], A. Calabro[34], A. Cappi[15,108], T. Castro[17,18,16,99], K. C. Chambers[109], S. Contarini[22], A. R. Cooray[110], S. Davini[25], B. De Caro[33], G. Desprez[111], A. Díaz-Sánchez[112], S. Di Domizio[24,25], H. Dole[11], S. Escoffier[67], A. G. Ferrari[41,21], I. Ferrero[59], F. Fornari[93], L. Gabarra[113], K. Ganga[75], J. García-Bellido[100], E. Gaztanaga[39,40,114], F. Giacomini[21], F. Gianotti[15], G. Gozaliasl[115,69], A. Hall[4], S. Hemmati[116], H. Hildebrandt[117], J. Hjorth[118], A. Jimenez Muñoz[119], J. J. E. Kajava[120,121], V. Kansal[122,123], D. Karagiannis[124,125], C. C. Kirkpatrick[66], J. Le Graet[67], L. Legrand[126], A. Loureiro[127,128], J. Macias-Perez[119], G. Maggio[17], M. Magliocchetti[50], F. Mannucci[129], R. Maoli[130,34], C. J. A. P. Martins[131,29], S. Matthew[4], L. Maurin[11], R. B. Metcalf[74,15], P. Monaco[106,17,18,16], C. Moretti[19,99,17,16,18], G. Morgante[15], Nicholas A. Walton[132], L. Patrizii[21], V. Popa[82], D. Potter[133], P. Reimberg[76], I. Risso[95], P.-F. Rocci[11], R. P. Rollins[4], M. Sahlén[134], A. Schneider[133], M. Sereno[15,21], P. Simon[2], A. Spurio Mancini[135,6], K. Tanidis[113], C. Tao[67], G. Testera[25], R. Teyssier[136], S. Toft[62,137,138], S. Tosi[24,25], A. Troja[84,51], M. Tucci[49], C. Valieri[21], J. Valiviita[69,70], D. Vergani[15], and G. Verza[139,140]

*(Affiliations can be found after the references)*

**ABSTRACT**

Data from the *Euclid* space telescope will enable cosmic shear measurements to be carried out with very small statistical errors, necessitating a corresponding level of systematic error control. A common approach to correct for shear biases involves calibrating shape measurement methods using image simulations with known input shear. Given their high resolution, galaxies observed with the *Hubble* Space Telescope (HST) can, in principle, be utilised to emulate *Euclid* observations of sheared galaxy images with realistic morphologies. In this work, we employ a `GalSim`-based testing environment to investigate whether uncertainties in the HST point spread function (PSF) model or in data processing techniques introduce significant biases in weak-lensing (WL) shear calibration. We used single Sérsic galaxy models to simulate both HST and *Euclid* observations. We then 'Euclidised' our HST simulations and compared the results with the directly simulated *Euclid*-like images. For this comparison, we utilised a moment-based shape measurement algorithm and galaxy model fits. Through the Euclidisation procedure, we effectively reduced the residual multiplicative biases in shear measurements to sub-percent levels. This achievement was made possible by employing either the native pixel scales of the instruments, utilising the Lanczos15 interpolation kernel, correcting for noise correlations, and ensuring consistent galaxy signal-to-noise ratios between simulation branches. Alternatively, a finer pixel scale can be employed alongside deeper HST data. However, the Euclidisation procedure requires further analysis on the impact of the correlated noise, to estimate calibration bias. We found that additive biases can be mitigated by applying a post-deconvolution isotropisation in the Euclidisation set-up. Additionally, we conducted an in-depth analysis of the accuracy of `TinyTim` HST PSF models using star fields observed in the F606W and F814W filters. We observe that F606W images exhibit a broader scatter in the recovered best-fit focus, compared to those in the F814W filter. Estimating the focus value for the F606W filter in lower stellar density regimes has allowed us to reveal significant statistical uncertainties.

## 1. Introduction

The light from distant galaxies travelling through the Universe is deflected due to the gravitational potential generated by the large-scale matter distribution. Typically, the fluctuations in the intervening mass distribution cause a slight, coherent distortion, which is imprinted onto the observed shapes of the galaxies, commonly referred to as weak gravitational lensing (WL, see e.g. Bartelmann & Schneider 2001 for a detailed introduction). The low amplitude of the WL signal makes probing the growth of structure in the Universe and deriving cosmological information technically challenging. Nonetheless, this can be achieved by measuring the shapes of a large number of galaxies. Measurements of the correlation between galaxy shapes, known as cosmic shear (see, e.g. Kilbinger 2015,Mandelbaum 2018 for reviews), can provide insights into the statistical properties of the mass distribution. Additionally, cosmic shear allows us to investigate the evolution of structure on large scales and explore the geometry of the Universe. Obtaining such measurements is a primary goal of current and future large dedicated surveys; for instance, the space missions of *Euclid*[1] (Laureijs et al. 2011) and the *Nancy Grace Roman* Space Telescope[2] (Spergel et al. 2015) as well as the ground-based *Vera C. Rubin* Observatory's Legacy Survey of Space and Time[3] (*Rubin*-LSST; Ivezić et al. 2019).

In this work, we focus on the *Euclid* mission, which will survey about 14 000 deg$^2$ of the sky in the optical and near-infrared (Euclid Collaboration: Mellier et al. 2024), aiming to obtain unprecedented WL constraints on the large-scale structure (LSS) of the Universe. *Euclid* is optimised for obtaining WL measurements thanks to its optimal conditions for accurate galaxy shape measurements. This is made possible thanks to the stable observing conditions and high spatial resolution achieved by being in space, as well as its design. The latter minimises any corrections for the blurring caused by the point-spread function (PSF), and the stability allows the PSF to be known accurately as a function of time, position, and wavelength across the field-of-view. *Euclid* has a wide field of view of 0.54 deg$^2$ with a broad optical band pass (VIS, see Euclid Collaboration: Cropper et al. 2024), covering approximately the range 530–920 nm, maximising the number of observed galaxies. However, the observations will still be compromised by some factors. For example, the PSF with which the observed galaxies are convolved depends on the wavelength and, thus, on the spectral energy distribution (SED) in the observed frame. Hence, an incorrect estimate of the wavelength-dependent model for the PSF and/or the galaxy SED may bias the shear estimates (Cypriano et al. 2010; Eriksen & Hoekstra 2018). In addition, the SED of a galaxy typically varies spatially, generating 'colour gradient' (CG) bias (Voigt et al. 2012; Semboloni et al. 2013; Er et al. 2018; Kamath et al. 2019). Furthermore, the bias depends on the width of the filter that is used (Semboloni et al. 2013). Consequently, CG bias is expected to be particularly relevant for *Euclid* because of its wide pass-band (Laureijs et al. 2011).

The *Hubble* Space Telescope (HST), with its high angular resolution and multiple filters covering the *Euclid* band pass, provides the most suitable data set to calibrate *Euclid* shear measurements against CG bias. We can use HST images of a representative sample of galaxies that *Euclid* will observe in order to accurately calibrate the shear measurement biases. Moreover, a sufficient number of galaxies has already been observed to calibrate the bias with the precision required to achieve *Euclid*'s science objectives (Semboloni et al. 2013).

Multi-band HST galaxy images, together with the supporting PSF models, can be used for *Euclid* calibrations via three approaches. In the first approach, models are fit to the galaxies providing distributions of galaxy parameters. These can be used as input distributions for image simulations based on parametric galaxy models (e.g. Hoekstra et al. 2017; Kannawadi et al. 2019; Hernández-Martín et al. 2020). In the second approach, HST postage stamps are directly used as input to the image simulations to render fully realistic morphologies (Mandelbaum et al. 2012; Mandelbaum et al. 2015; Rowe et al. 2015; Bergamini et al., in preparation). Finally, generative machine learning models can be used to render *Euclid*-like images (Lanusse et al. 2021).

In this paper, we focus on the second approach and analyse a procedure called 'Euclidisation' which involves using simulated HST galaxy images to create emulated *Euclid* observations of galaxies with a known artificial WL shear. This procedure includes: deconvolving the emulated HST galaxy images by the HST PSF, adding shear, convolving with the *Euclid* PSF, adjusting pixel noise, flux scaling, resampling, and adding further noise (see Fig. 1, bottom branch). To test the procedure and quantify the impact of uncertainties in the HST data (in particular, regarding the PSF model), we applied the Euclidisation procedure to simulated input galaxies. For these we vary properties such as half-light radius, Sérsic index, and signal-to-noise ratio (S/N).

⋆ e-mail: dianas@jpl.nasa.gov

[1] https://sci.esa.int/Euclid/

[2] https://roman.gsfc.nasa.gov/

[3] https://www.lsst.org/lsst

The resulting 'Euclidised' galaxy images are compared with directly emulated 'direct *Euclid*-like' images (see Fig. 1, top) in a testing environment. For both branches seen in Fig. 1, we could then compare shear measurement biases, which should be identical if the Euclidisation procedure emulates the sheared galaxy images correctly. Given the importance of the HST PSF model for the Euclidisation procedure, we also carried out an in-depth analysis of the accuracy of `TinyTim` PSF models for HST/ACS, using star fields observed in filters F606W and F814W, thereby extending the work done in Gillis et al. (2020). Furthermore, we investigated the accuracy of recovering the HST focus in a regime of low stellar density.

This paper is structured as follows. In Sect. 2, we discuss the formalism of the WL shear measurement biases assumed in our analysis. In Sect. 3, we describe the testing environment. The two methods for the galaxy shape and properties parameters measurements are described in Sect. 4. We present the different tests and findings in Sect. 5. In Sect. 6, we describe our investigation of the accuracy of the `TinyTim` PSF models for HST. We then summarise our results and discuss their significance in Sect. 7.

## 2. Shear measurement formalism

In the WL limit ($\kappa \ll 1$ and $|\gamma| \ll 1$, with $\kappa$ being the convergence and $\gamma$ the shear), for a given galaxy of intrinsic ellipticity $\epsilon^{\rm int}$ having undergone a lensing-induced shear $\gamma$, the observed ellipticity is the sum of the intrinsic ellipticity and the shear, expressed as

$$\epsilon_i^{\rm obs} \approx \epsilon_i^{\rm int} + g_i\,, \qquad i = 1,2\,, \tag{1}$$

where $g = \gamma/(1+\kappa)$ is the reduced shear, which in the following we refer to as shear for simplicity, and $i$ identifies the two components of shear. Assuming random intrinsic ellipticity orientations, the expectation value is $\left\langle \epsilon^{\rm obs} \right\rangle = g$, since $\left\langle \epsilon^{\rm int} \right\rangle = 0$. The dispersion of the observed ellipticity is $\sigma\left(\epsilon^{\rm obs}\right) \simeq \sqrt{\sigma^2\left(\epsilon^{\rm int}\right) + \sigma_{\rm m}^2}$, which has contributions from both the intrinsic ellipticity dispersion $\sigma\left(\epsilon^{\rm int}\right) = \sigma\left(\epsilon^{\rm obs} - g\right)$ of the galaxy sample and measurement root mean square (RMS) errors, $\sigma_{\rm m}$ (e.g. Hoekstra et al. 2000, Leauthaud et al. 2007, Schrabback et al. 2018). The intrinsic ellipticity dispersion is the dominant term by an order of magnitude.

Systematic errors affect the measurement of galaxy ellipticity. To control these systematic errors, any shear measurement method needs to be calibrated through simulations to quantify possible differences between the input and the recovered shear. Analyses distinguish between the additive bias $c_i$ with $i = 1,2$ which adds a value $c_i$ to the true shear, and the multiplicative bias $\mu_i$ with $i = 1,2$[4], which distorts the amplitude of the shear by a factor of $(1+\mu_i)$. Additive bias can be caused, for instance, by an insufficient correction of the PSF anisotropy and may lead to spurious correlations in the shape of galaxies (Massey et al. 2013). Conversely, dominant sources for multiplicative bias include noise bias (Refregier et al. 2012), model bias (e.g. Melchior & Viola 2012; Refregier et al. 2012; Miller et al. 2013; Kacprzak et al. 2014), or the impact of neighbouring objects (Hoekstra et al. 2017, 2021; Euclid Collaboration: Martinet et al. 2019). Thus, knowing the input shear $g^{\rm true}$ in WL image simulations, at the first order, one usually fits the recovered shear $g^{\rm obs}$ as

$$g_i^{\rm obs} = (1+\mu_i)\, g_i^{\rm true} + c_i, \qquad i = 1,2\,, \tag{2}$$

for both components of the shear separately. The typical change in ellipticity caused by cosmic shear is about one per cent, which is much smaller than the intrinsic ellipticities of galaxies and also smaller than the typical biases introduced by instrumental effects. In this work, we only investigate linear-order biases, but we note that *Euclid* also has strict requirements on quadratic biases. We refer to Kitching & Deshpande (2022) for more details.

## 3. Simulation and analysis set-up

To quantify the uncertainties in using HST images as input for WL image simulations, we built a testing environment. It employs simulated data that approximately resemble the properties of HST/ACS and *Euclid*/VIS observations (see detailed description in Sect. 5). This section describes the methodology and key components of our set-up, including the testing environment and simulation size, while addressing the effects of shape cancellation noise.

### 3.1. Testing environment

We generated galaxy image simulations with the open source `GalSim` software[5] (Rowe et al. 2015). The following outlines the steps of the testing environment featured in Fig. 1.
Input galaxy. As input for our simulations, we model the galaxy light profile with a single component Sérsic model (Sérsic 1963) as

$$I(R) = I_{\rm e} \exp\left[-b_n\left((R/R_{\rm e})^{1/n} - 1\right)\right]\,, \tag{3}$$

with the half-light radius, $R_{\rm e}$, which is the radius containing half of the total luminosity of the galaxy, the intensity at that radius, $I_{\rm e}$, and the parameter $b_n \approx 2n - 1/3$, where $n$ is the Sérsic index. Equation (3) can be also written as

$$I(x) = A\,\exp\left(-k\left[(\mathbf{x}-\mathbf{x_0})^T\, C^{-1}\,(\mathbf{x}-\mathbf{x_0})\right]^{1/(2n)}\right)\,, \tag{4}$$

where $\mathbf{x_0}$ is galaxy centroid, $A$ is the peak intensity, $C$ is the galaxy covariance matrix, which includes elements associated with ellipticity (see e.g. Voigt & Bridle 2010 for more details), $k = 1.9992n - 0.3271$, and $n$ is the Sérsic index. It is worth noting that, although this model is less realistic than two-component models, it is a reasonably realistic model for which the intrinsic ellipticity is well-defined, and this has some significant advantages in analysing results.

We created postage stamp images of isolated galaxies of a size of $512 \times 512$ pixels, with a pixel scale of 0″.02 or 0″.04. The choice depends on the specific test (see Sect. 5 for different test scenarios). The large postage stamp size prevents issues related to the dilation of the galaxies during subsequent steps of the procedure. We also tested different interpolation kernels to study their impact on the bias results (discussed in Sects. 5.2 and 5.5). Each mock galaxy was initially arbitrarily assign a flux of 10 000 ADU. Later, this flux will be rescaled according to the properties of the HST and *Euclid* telescopes, to obtain four different values for the S/N, from 10 to 40 with a step size of 10. These values were computed using the CCD equation (Howell 1989) as detailed in Appendix A. To probe the sensitivity of our analysis to intrinsic galaxy properties, we conducted the analysis across a range of half-light radii, $R_e[''] \in \{0.2, 0.3, 0.4, 0.5, 0.6, 0.7\}$, and Sérsic indices of $n \in \{1, 2, 3\}$. These parameters were drawn from a random uniform distribution within these ranges, ensuring an even sampling of galaxy sizes and Sérsic profiles.

[4] We treat the multiplicative bias $\mu$ as a spin-2 quantity, scaling $g_1$ and $g_2$ independently, assuming no cross-component effects ($\mu_{12} = \mu_{21} = 0$).

[5] https://github.com/GalSim-developers/GalSim

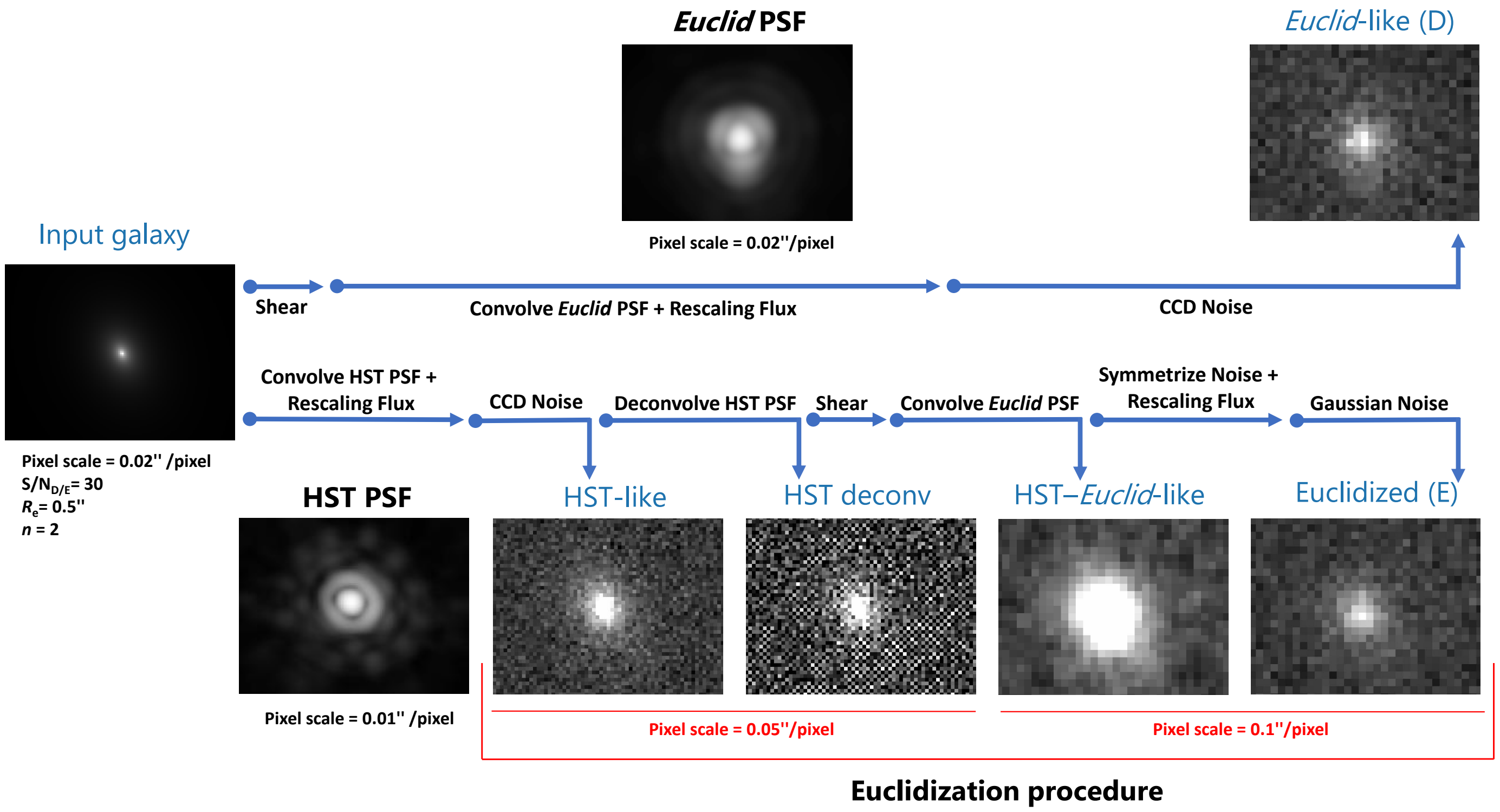

**Fig. 1.** Illustration of the testing environment to create direct *Euclid*-like (D) images to be compared with Euclidised (E) images. See the text in Sect. 3 for details.

We assigned the intrinsic ellipticity components $\epsilon_1$ and $\epsilon_2$ drawn from a Gaussian distribution with zero mean and dispersion $\sigma = 0.3$ to the galaxies; however, we excluded galaxies with very high ellipticities $|\epsilon| > 0.7$. We then applied a random shift to the galaxy position with a uniform distribution from $-0\farcs05$ to $0\farcs05$ in both axes to have a small random displacement with respect to the pixel centre (e.g. half the fiducial *Euclid* pixel scale), as is the case for real data. Then, for each galaxy, we created a second galaxy, which is identical but orthogonally oriented, to mitigate the intrinsic shape noise (Nakajima & Bernstein 2007; Massey et al. 2007; Mandelbaum et al. 2014).

At this point, two versions of the same mock galaxy were drawn to be simulated as 'direct *Euclid*-like observations' (hereafter, D) and Euclidised image (hereafter, E), as shown in Fig. 1 (top and bottom, respectively). The following steps detail this process.

Direct *Euclid*-like. The input galaxy pair is sheared by a value taken from a discrete uniform distribution in the range from $-0.06$ to $0.06$ with a step of 0.004, using the `GalSim` function `galsim.lens` without magnification. Sheared galaxy images are then convolved with the *Euclid*-like PSF. We employed a model of the VIS PSF, which was computed with the `PSFToolkit` (Duncan et al., in preparation) from a realistic simulation of galaxy SEDs using the Empirical Galaxy Generator (EGG; Schreiber et al. 2017), at the centre of the field of view, assuming a physical model of the telescope. This PSF model is sampled on a grid of $0\farcs02$ pixel$^{-1}$, which is five times finer than the native VIS pixel grid. The flux of the galaxy is then empirically rescaled so that its measured S/N (see Appendix A) statistically reaches the desired value, given the simulated *Euclid* conditions. We applied noise in our simulations using the `GalSim` function `CCDNoise`. It includes Poisson shot noise from the source and the sky background, and Gaussian read-out noise. A new random seed was drawn for each pair of orthogonally oriented galaxies to match the noise between them. We assumed a sky brightness of $m_{\rm sky} = 22.35$ mag in one square arcsecond (Refregier et al. 2010). Following Tewes et al. (2019), we can compute the corresponding sky level as

$$F_{\rm sky}\,[\mathrm{ADU}] \;=\; \frac{t_{\rm exp}\,[\mathrm{s}]}{\mathrm{gain}\,[\mathrm{e}^-/\mathrm{ADU}]}\,10^{-0.4\left(m_{\rm sky}-\mathrm{zp}\right)}\left(l\,['']\right)^2\,, \tag{5}$$

where we assume an exposure time $t_{\rm exp}$ for *Euclid* of 1695 s, corresponding to the co-addition of three single exposures of 565 s each (Laureijs et al. 2011). While *Euclid* typically takes four exposures at each pointing position, a large fraction of the survey will only be covered by three exposures due to chip gaps (Euclid Collaboration: Scaramella et al. 2022), justifying this assumption. We assume a CCD gain of 3.1 $\mathrm{e}^-\,\mathrm{ADU}^{-1}$ (Niemi et al. 2015), a read-out noise (see Appendix A) of 4.2 $\mathrm{e}^-$ (Cropper et al. 2016). We adopted an instrumental zero-point of 24.6 mag (Tewes et al. 2019) and a pixel size of $l = 0\farcs1$ (Laureijs et al. 2011). Once the noise is applied, we can obtain the 'direct *Euclid*-like' image (D).

The bottom branch of the diagram in Fig. 1 illustrates how the Euclidised image was obtained:

HST-like. The same input galaxy pair is then convolved with the HST PSF created with `TinyTim` (see Sect. 6 and Appendix B

for a detailed analysis). The flux is rescaled to take into account the properties of the HST and, finally, CCD noise is added in order to create HST-like images. For the HST observations, we assumed an exposure time of 1000 s and a S/N that is twice the value of the direct *Euclid*-like galaxy to represent the difference in the mirror size between the two telescopes. We adopted a CCD gain of 2.0 $e^-$/ADU, a read-out noise of 5.0 $e^-$. We assumed a sky background in one square arcsecond to have an average value of $m_{\rm sky}$ = 22.5 mag and a zero-point of 25.9 mag for F814W.[6] We also simulated HST observations in the F606W filter, with a zero-point of 26.5 mag.

HST deconv. We then performed a 'reconvolution' process. as described in Rowe et al. (2015). We deconvolved the HST-like images by the HST PSF using the `GalSim` class `galsim.Deconvolution`, which is based on a division in Fourier space. To analyse the impact of HST PSF model uncertainties, we may use a different PSF for this deconvolution step than for the prior convolution (see Sect. 5.5 for details).

HST–*Euclid*-like. We added shear to the HST deconv images and convolved them with the *Euclid* PSF.

Euclidised. The images resulting from the convolution with the *Euclid* PSF carry correlated noise. The isotropisation (or symmetrisation) of this noise enforces a four-fold symmetry, introducing minimal extra noise through the GalSim function `symmetrizeImage`. Then, after rescaling the flux, some extra Gaussian noise with dispersion, $\sigma_{\rm G}$, is added to the stamp in order to match the noise level of the "direct" branch. For simplicity, in this step we did not add further Poisson noise from the photon counts of the sources. However, we did include Poisson noise when we generated *HST-like* images.

At this point, we obtained the Euclidised image, E, which can be compared to the direct *Euclid*-like images, D, and analysed in exactly the same manner. Our test procedure does not include detection and deblending steps, as we are simulating images of isolated galaxies. As a result of this simplification, our study does not suffer from the object detection bias discussed in Sheldon et al. (2020) and Hoekstra et al. (2021). In the scope of this paper, we only tested the Euclidisation of simple Sérsic profile galaxies.

### 3.2. Simulation size and shape noise cancellation

To match the statistical precision of *Euclid*, systematic shear measurement biases will need to be controlled to an accuracy of $|\delta\mu| < 2 \times 10^{-4}$ and $|\delta c| < 5 \times 10^{-5}$ (Cropper et al. 2013). For this purpose, the sources of statistical uncertainty can be constrained by averaging over large numbers of galaxies given by (e.g. Mandelbaum 2018, Fenech Conti et al. 2017):

$$N_{\rm gal} = \left(\frac{\sigma_\epsilon}{|\delta\mu||g|}\right)^2 , \tag{6}$$

where $\sigma_\epsilon$ = 0.3 is the dispersion of galaxy ellipticities and $|g|$ is the modulus of the shear applied to our simulations. For a shear modulus of 0.03 on average, in principle, we need $2.5\times10^9$ galaxies to constrain the multiplicative bias to $|\delta\mu| < 2 \times 10^{-4}$.

In order to reduce the required simulation size, we employed shape noise cancellation (see Sect. 3), which we found empirically to reduce the sample size by a factor of approximately 4, under the specific conditions and parameter settings of our simulation, compared to Eq. (6). We note that for our testing environment shape noise cancellation not only reduces the required simulation volume, but also minimises the impact that correlations caused by the identical intrinsic shapes in the D and the E images have on the bias analysis.[7]

We note that further approaches have been proposed to reduce simulation volume in addition to shape noise cancellation, such as measurements of the shear response for individual galaxies (Pujol et al. 2018) and pixel noise cancellation (Euclid Collaboration: Martinet et al. 2019). We refer to Jansen et al. (2024) for a comparison of the efficiencies of these different approaches.

For our simulation, we drew about $10^7$ galaxy stamps for each setting (including the rotated galaxies), which is sufficient to reach a precision on the multiplicative bias of about $10^{-3}$ (see Sect. 5). Moreover, this allows us to recover a meaningful correction approximately at the level of the *Euclid* requirements.

## 4. Galaxy property measurements

In this section, we describe our galaxy shape and parameter measurements on the direct *Euclid*-like and Euclidised images to check the accuracy of the Euclidisation. As comparison metrics, we use two approaches: measuring the biases in the shear recovery using the moment-based KSB galaxy shape measurements (see Sect. 4.1; also Kaiser et al. 1995; Luppino & Kaiser 1997; Hoekstra et al. 1998) and estimating the galaxy parameters by fitting a galaxy model (see Sect. 4.2).

### 4.1. KSB measurements

The galaxy shapes are measured using the `galsim.hsm.EstimateShear`[8] function with `KSB` as desired method for PSF correction. This implementation requires the PSF and galaxy images to have the same pixel scale. However, the *Euclid* PSF is actually created with a pixel scale of 0\farcs02, namely, it is over-sampled by a factor of 5 with respect to the native pixel scale, so as to avoid losing relevant details. To use this method, in the tests described in Sect. 5, we chose either to over-sample the galaxy image to 0\farcs02 or use a pixel scale of 0\farcs04 (i.e. matching the HST/UVIS pixel scale) for each step of the procedure, depending on what we want to investigate. When we over-sample, the estimates of the second moments of an image are evaluated on a finer grid, but the intensity is taken constant over the sub-pixels and not interpolated. Furthermore, in the cases where we over-sample the galaxy images from the *Euclid* pixel scale of 0\farcs1 to 0\farcs02 before running KSB, we also convolve the *Euclid* PSF model with a 2D top-hat profile of 0\farcs1 × 0\farcs1. This is because observations done with large pixels lead to a loss of resolution. If we artificially over-sample an image with pixels 0\farcs1 to 0\farcs02 we do not recover that loss in resolution. When applying the KSB method, the specified PSF must contain all the convolutive effects that were applied to the galaxy image after being sheared. This includes the loss of resolution from diffraction by the telescope optics (which is captured by an over-sampled PSF model), but also the loss of resolution due to the pixellation by the detector array, in the present case not captured by the over-sampled PSF model. One way to take this into account is to include a pixel convolution before applying the KSB method, thus avoiding propagating errors to the KSB shape measurements.

[6] This information is available on the "Advanced Camera for Surveys Instrument Handbook" (version 9.0) from `https://www.stsci.edu/itt/review/ihb_cy15/ACS/ACS_ihb.pdf`

[7] We cross-checked whether the remaining correlation has an impact by estimating the difference of the measured shears between the two outputs directly. This led to results that are consistent within the errors with what we obtain when fitting the two outputs separately and then calculating the bias differences (see Sect. 4.1).

[8] The `GalSim` `KSB` algorithm is a specific implementation of the KSB method (Kaiser et al. 1995; Luppino & Kaiser 1997), as described in Appendix C of Hirata & Seljak (2003).

The KSB method measures the moments of the surface-brightness distribution of stars and galaxies to infer PSF-corrected estimates of galaxy ellipticities. It parametrises galaxies according to their weighted quadrupole moments and describes the PSF as a small but highly anisotropic distortion convolved with a large circularly symmetric function. Furthermore, all the tests we present in Sect. 5 use unit shear weights. With these assumptions, the KSB method returns a per-object estimate of the shear components $\hat{\epsilon}_1$ and $\hat{\epsilon}_2$.

We used the weighted least square (WLS) fit of the model described in Eq. (2) to measure the shear bias. The weights were determined as the reciprocals of the shear variance. We recovered the multiplicative bias term, $\mu_i^j$, and additive bias term, $c_i^j$, for both components, $i \in \{1, 2\}$, of the shear and both images $j \in \{\mathrm{E}, \mathrm{D}\}$, as the slope and the intercept of the fitting between the ellipticity and the input shear and their corresponding standard deviation (SD). We also calculate the differences $\mu_i^{\mathrm{D}} - \mu_i^{\mathrm{E}}$ and $c_i^{\mathrm{D}} - c_i^{\mathrm{E}}$, and we adopt the standard error of the mean as the error, using Gaussian error propagation.

The KSB method is computationally fast but, in some cases, its implementation fails to compute the shapes, or returns ellipticity estimates with an absolute value larger than 1. This occurs because the algorithm is not sufficiently robust when handling highly elliptical or small galaxies, resulting in situations where the iterative process fails to converge to a solution. In our analysis, this occurs especially at lower S/N or for large input ellipticities $|\epsilon| \gtrsim 0.7$. For this reason, we reject galaxies with estimates $|\hat{\epsilon}| > 1$. For the different tests we performed, this results in the removal of a small fraction of galaxies from our initial sample, see Table 1. For more details, we refer to Sect. 5. It is worth noting that the objective of this study is not to obtain a tight absolute calibration of this algorithm. We rather want to test the Euclidisation procedure by investigating the relative bias difference between the two branches and estimate the correction to apply to real data in shear measurement analyses.

### 4.2. Galaxy model fit

To obtain an alternative comparison metric, we fit two-dimensional elliptical Sérsic models to the PSF-convolved output galaxy images D and E. Employing the Astropy `EllipSersic2D` model,[9] galaxies are modelled with a single Sérsic profile, where the centroid position, the ellipticity, the total flux, $F$, the half-light radius, $R_\mathrm{e}$, and the Sérsic index, $n$, are estimated directly from the galaxy postage stamp. We note that this model intentionally ignores the PSF. The purpose of this fit is solely to compare the observed shape of galaxies as simulated in the D and E images.

We set the postage stamp dimensions to $512 \times 512$ pixels, a large enough image size to include the flux without encountering any image edge effects in the measurements of galaxy properties. However, the fit to estimate the galaxy parameters was performed using the Levenberg-Marquardt algorithm (LMA) within a smaller region, computed following the procedure described in Appendix A, containing an elliptical aperture that extends out to three half-light radii of the galaxy. This saves computational time because the effective postage stamp will be much smaller than the original $512 \times 512$ pixels.

In our analysis, we discarded unreliable fit results (e.g. no convergence or a Sérsic index $n$ outside the range $[0.1, 6.0]$). To further reduce the run-time, we performed the fit only on a subsample of galaxies and in some configurations of our pipeline. This is discussed further in the next section.

[9] We adjusted the default major and minor axis as follows: $a = R_\mathrm{e}/\sqrt{(1-g)/(1+g)}$ and $b = R_\mathrm{e}\sqrt{(1-g)/(1+g)}$ in order to have a match between the different definitions in Astropy and `GalSim`.

## 5. Tests and results

In this section we evaluate the accuracy of our Euclidisation procedure (and its variants) under different conditions. Our goal is to minimise the potential impact of the Euclidisation procedure on shear bias. An overview of the different tests conducted is provided in Table 1.

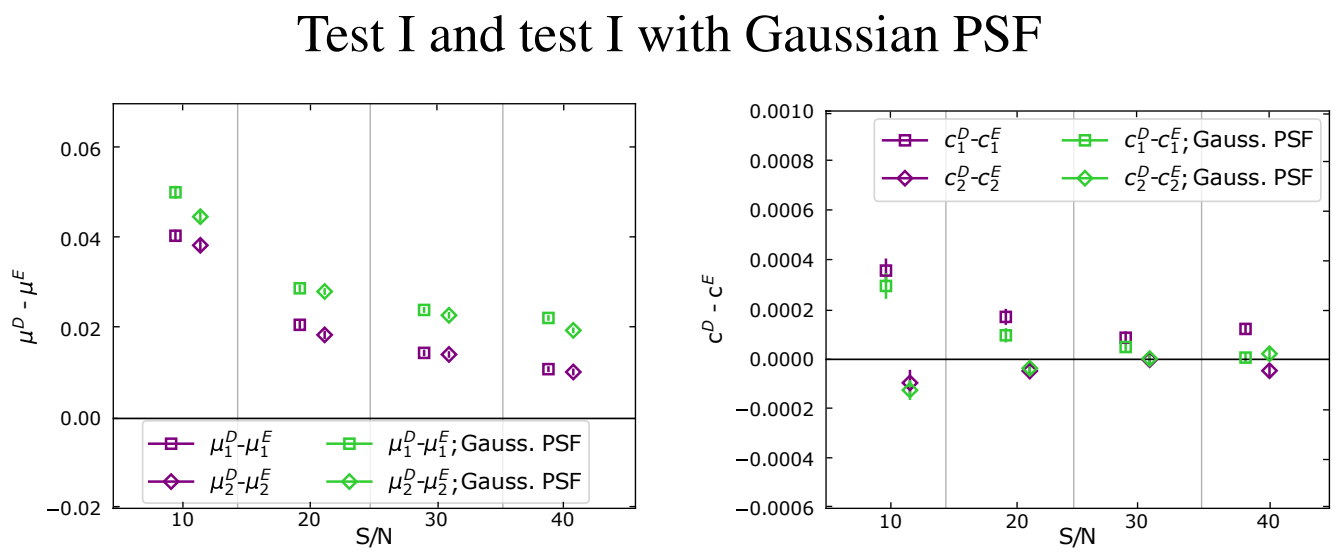

**Fig. 2.** Comparing bias estimates when using a *Euclid* PSF versus a Gaussian PSF. Multiplicative (left panel) and additive (right panel) shear bias differences obtained employing a *Euclid*/VIS pixel scale of 0."1 and a `quintic` interpolation kernel, as described in Sect. 5.1, without the correction for the impact of noise correlations on S/N estimates described in Sect. 5.2. The data points show the bias obtained from the fit with the error bars indicating 1 $\sigma$ uncertainties for a number of galaxies $N_\mathrm{gal} \simeq 24.1 \times 10^6$ for each S/N.

### 5.1. Use of native HST/ACS and Euclid/VIS pixel scales

In the first test, referred to as 'test I', starting with an *input galaxy* with a pixel scale of 0."02. We employ the testing environment using the native pixel scales of 0."05 and 0."1 for the simulated HST/ACS and *Euclid*/VIS images, respectively. This approach is illustrated in Fig. 1. Hence, the two output images D and E were drawn with the native *Euclid*/VIS pixel scale of 0."1. However, for the computation of galaxy shapes using the KSB method, the two galaxies were over-sampled by a factor of 5 to compute the KSB moments consistently from the PSF image and galaxy postage stamp with a fine pixel scale of 0."02. We used the `quintic` image interpolation scheme to specify how interpolation should be done at locations in between the integer pixel centres, which is the default option in `GalSim`. The number of galaxies, $N_\mathrm{gal}$, for each sample of D and E galaxies and for each S/N we employed in our tests, the KSB failure rates, and the interpolation kernel are reported in Table 1.

Figure 2 (purple symbols) shows the multiplicative and additive bias differences ($\Delta\mu$ and $\Delta c$) between the two outputs, for four different values of S/N. The data points show the bias obtained from the fit with error bars indicating the 1 $\sigma$ standard

**Table 1.** Tests presented in Sect. 5.

| Test | $N_{\rm gal}$ $[10^6]$ | Failure rate [%] D | E | Pixel scale [″] D/E | Interpolation kernel | Description |
|---|---|---|---|---|---|---|
| I | 24.1 | 4.9 | 4.6 | 0.1 | `quintic` | HST/ACS and *Euclid* native pixel scales |
| I with Gauss. PSF | 24.1 | 4.4 | 4.1 | 0.1 | `quintic` | Similar to I, but with Gaussian PSF |
| I with $10 \times {\rm S/N_{HST}}$ | 37.0 | 4.4 | 4.0 | 0.1 | `quintic` | Similar to I, but with higher $\rm S/N_{HST}$ |
| Ic | 1.4 | 4.9 | 3.4 | 0.1 | `lanczos15` | Similar to I, but with correction for noise correlation |
| Ic with $\sigma_{\rm G} = 2.5$ | 8.9 | 4.9 | 3.4 | 0.1 | `quintic` | Testing sensitivity to noise scaling with $\sigma_G = 2.5$ |
| Ic with $\sigma_{\rm G} = 2.7$ | 8.9 | 4.9 | 3.4 | 0.1 | `quintic` | Testing sensitivity to noise scaling with $\sigma_G = 2.7$ |
| Ic with PSF stack | 18.3 | 4.9 | 4.7 | 0.1 | `lanczos15` | Similar to Test Ic, but with HST average star stack and average model stack PSFs |
| II | 33.8 | 5.9 | 6.4 | 0.04 | `quintic` | Similar to I but, testing finer pixel scale |
| III | 24.7 | 4.7 | 5.1 | 0.04 | `quintic` | Similar to Test II, but using HST PSFs with different centre positions and focus |
| III with rotat. | 24.7 | 4.7 | 5.1 | 0.04 | `quintic` | Similar to Test III, but adding 'post-deconvolution isotropisation' of HST images |
| IV with $n = 1$ | 77.0 | 5.6 | 5.8 | 0.04 | `quintic` | Similar to II but, testing the impact of $R_{\rm trunc}$ for galaxies with $n = 1$ |
| IV with $n = 2$ | 77.0 | 2.2 | 2.2 | 0.04 | `quintic` | Similar to IV with $n = 1$, but for galaxies with $n = 2$ |
| IV with $n = 3$ | 77.0 | 1.6 | 1.5 | 0.04 | `quintic` | Similar to IV with $n = 1$, but for galaxies with $n = 3$ |

**Notes.** We report the number of galaxies $N_{\rm gal}$ for each S/N, the average failure rate of the KSB method over the four S/N levels, the pixel scales, the interpolation kernel used for both output images D and E, and a brief description of each test.

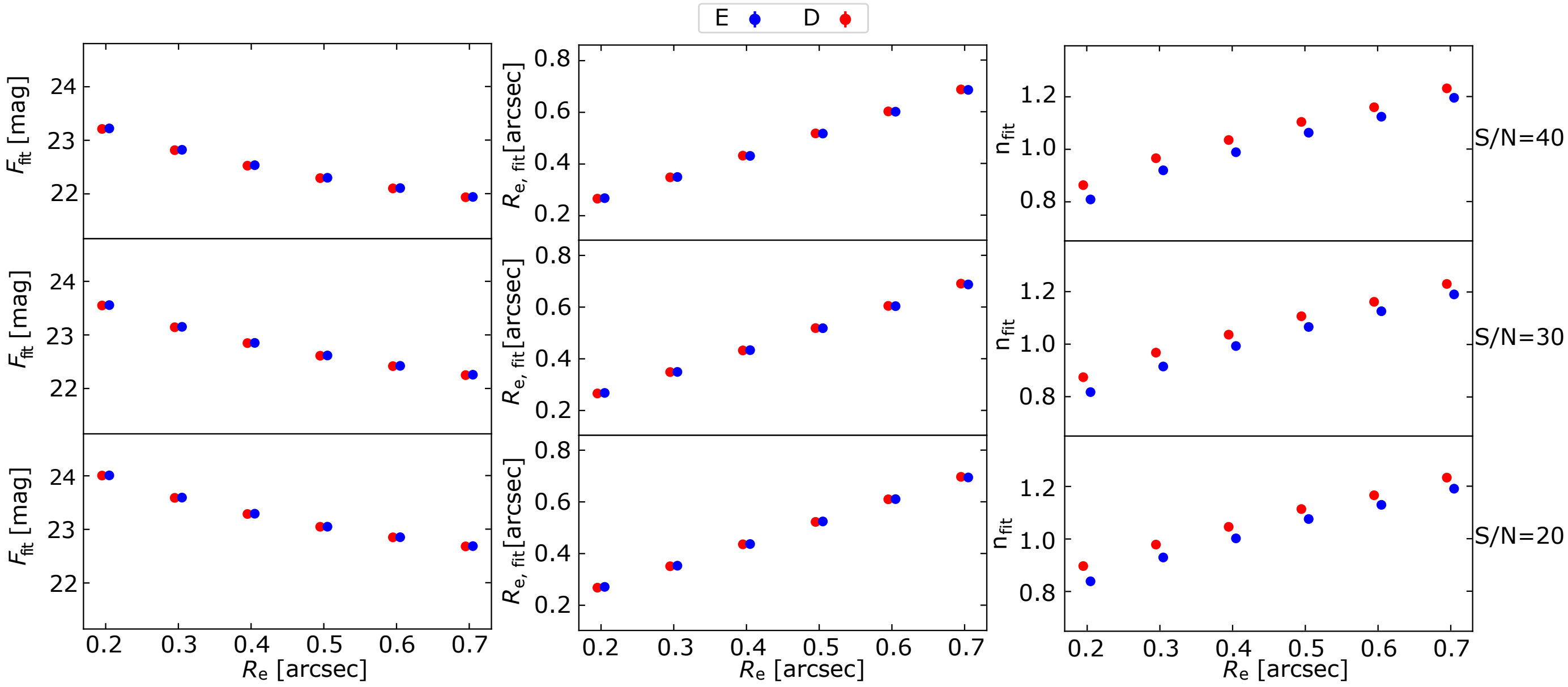

**Fig. 3.** Results from the galaxy model fits for test I using the native HST/ACS and *Euclid*/VIS pixel scales, for a number of galaxies $N_{\rm gal} \simeq 10^4$, as described in Sect. 5.1. The fitted parameters $R_{\rm e,fit}$ [″], $F_{\rm fit}$ [mag], and $n_{\rm fit}$ are shown as a function of the input $R_{\rm e}$ in arcsec for both the output galaxies D and E and for three values of the S/N. The data points are the average over the number of galaxies in each sample. The error bars, representing the $1\,\sigma$ uncertainty on the mean, are smaller than the size of the points and, thus, they are not visible.

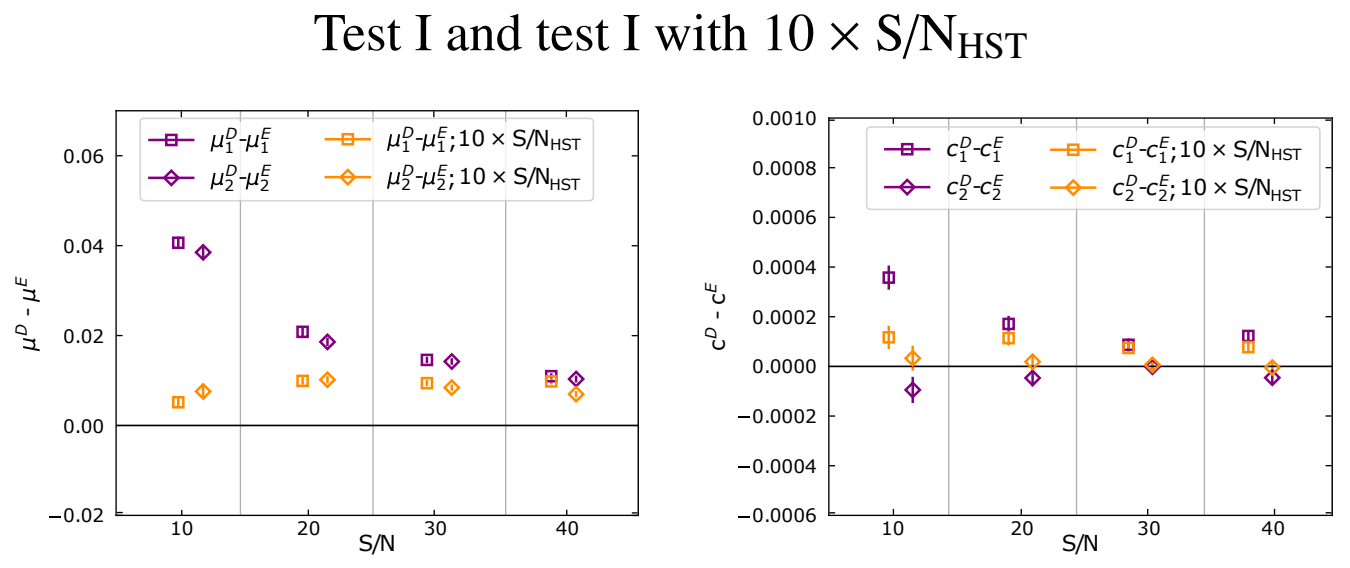

**Fig. 4.** Impact of using deeper HST/ACS images. Multiplicative (left panel) and additive (right panel) shear bias differences obtained employing a *Euclid*/VIS pixel scale of 0.″1, a `quintic` interpolation kernel (test I), and increasing the S/N of the HST-like images by a factor of 10 (test I with $10 \times {\rm S/N_{HST}}$), as described in Sect. 5.1. The data points show the bias obtained from the fit with the error bars indicating $1\,\sigma$ uncertainties for a number of galaxies $N_{\rm gal} \simeq 37.0 \times 10^6$ for each S/N.

error of the mean. The $\Delta\mu$ is larger for small galaxy S/N, evolving from 0.040 at S/N = 10 to 0.0107 at S/N = 40, averaging over both shear components. While the difference of the additive biases for the second component is consistent with zero for some of the S/N values, the first component decreases toward zero only at higher S/N, minimally increasing again at S/N = 40. The difference in additive bias components is primarily driven by differences in the ellipticity components of the *Euclid* PSF model. We expect that the observed bias difference depends on the shape measurement method (Euclid Collaboration: Congedo et al. 2024). Therefore, it is essential to establish a correction based on the specific shape measurement method utilised in the actual *Euclid* analysis. In general, the $\Delta c$ is of the order of $10^{-4}$. This is also the case for some of the other scenarios we explore. Therefore, we will focus on discussing $\Delta c$ only when we detect notable variations in additive bias differences, indicating deviations from the typical order of $10^{-4}$.

Clearly, the significant percent-level bias in the $\Delta\mu$ differences obtained with this first configuration are beyond any accuracy requirements on the Euclidisation procedure. To test if the *Euclid* PSF shape (but not the size) can have an impact on the bias difference, we repeat the analysis with a circular Gaussian PSF. The width of this Gaussian is set to $\sigma = 0''.07$, corresponding to the best fit to the detailed *Euclid* PSF. This Gaussian PSF is sampled with a pixel scale of $0''.02$, as before. The results for 'test I with a Gaussian PSF' are shown in Fig. 2 as well. The multiplicative bias difference is slightly increased, by 0.01 on average over the S/N. There is no significant difference for the additive bias, which remains consistent within the error with the default set-up for most of the S/N bins.

In Fig. 3, we present the results of the fit of the galaxy model, namely: the half-light radius $R_{\rm e,fit}$, the flux $F_{\rm fit}$, and the Seŕsic index $n_{\rm fit}$, as functions of the input half-light radius, for three different values of the S/N. We find a good agreement between D and E for the half-light radius and the flux. This is not the case for $n_{\rm fit}$, for which we obtain slightly lower estimates in E than D, consistently over all $R_{\rm e}$. This suggests that the Euclidised galaxy images, E, are slightly less centrally peaked than the direct *Euclid*-like images, D, likely contributing to the multiplicative bias difference. We note that it is not surprising that the recovered $n_{\rm fit}$ are generally lower than the input $n$ given that the fitted model does not correct for the smoothing impact of the PSF.

To identify the origin of the multiplicative bias difference (see Fig. 2) and of the shift in the recovered Sérsic index, we then tested whether these discrepancies are related to the noise level of the HST images, as part of 'test I with $10 \times {\rm S/N_{HST}}$'. We increased the S/N of the emulated HST galaxy images by a factor of 10, representing the use of much deeper HST data for the Euclidisation.[10] This increase in S/N leads to a reduction of the (averaged) bias difference compared to test I, as shown in Fig. 4. The multiplicative bias difference does not vanish completely, but it remains approximately constant for all the S/N. In addition, as the following sections show, an increase of HST depth is not a guarantee for a bias-free Euclidisation. The additive bias difference is below $2 \times 10^{-4}$ for both components.

### 5.2. Correcting the S/N for the impact of correlated noise

The noise model for a CCD image is typically a combination of Poisson noise on the pixel counts and Gaussian read noise. In WL measurements, this noise is commonly modelled as stationary on the scale of galaxy images, with the same variance for each pixel. This holds true when the Poisson noise on the sky level dominates and when the sky level does not vary much across each galaxy. In an idealised unprocessed image, the noise is largely uncorrelated between pixels, with the dispersion of the sum over $N$ pixel values scaling as the dispersion computed from single pixel values multiplied by $\sqrt{N}$. However, significant correlations between the noise in different pixels can be induced via processes such as correction for charge transfer inefficiency (CTI; Massey 2010), convolutions (Hartlap et al. 2009), and image resampling (Fruchter 2011; Rowe et al. 2011), particularly if the images are resampled to smaller pixels than those on the original detector (Gurvich & Mandelbaum 2016). In the presence of correlated noise, a naive estimation of noise based on the dispersion of single pixel values would underestimate the amount of noise in an aperture of interest (Casertano et al. 2000). This would lead to an overestimation of the S/N of a source.

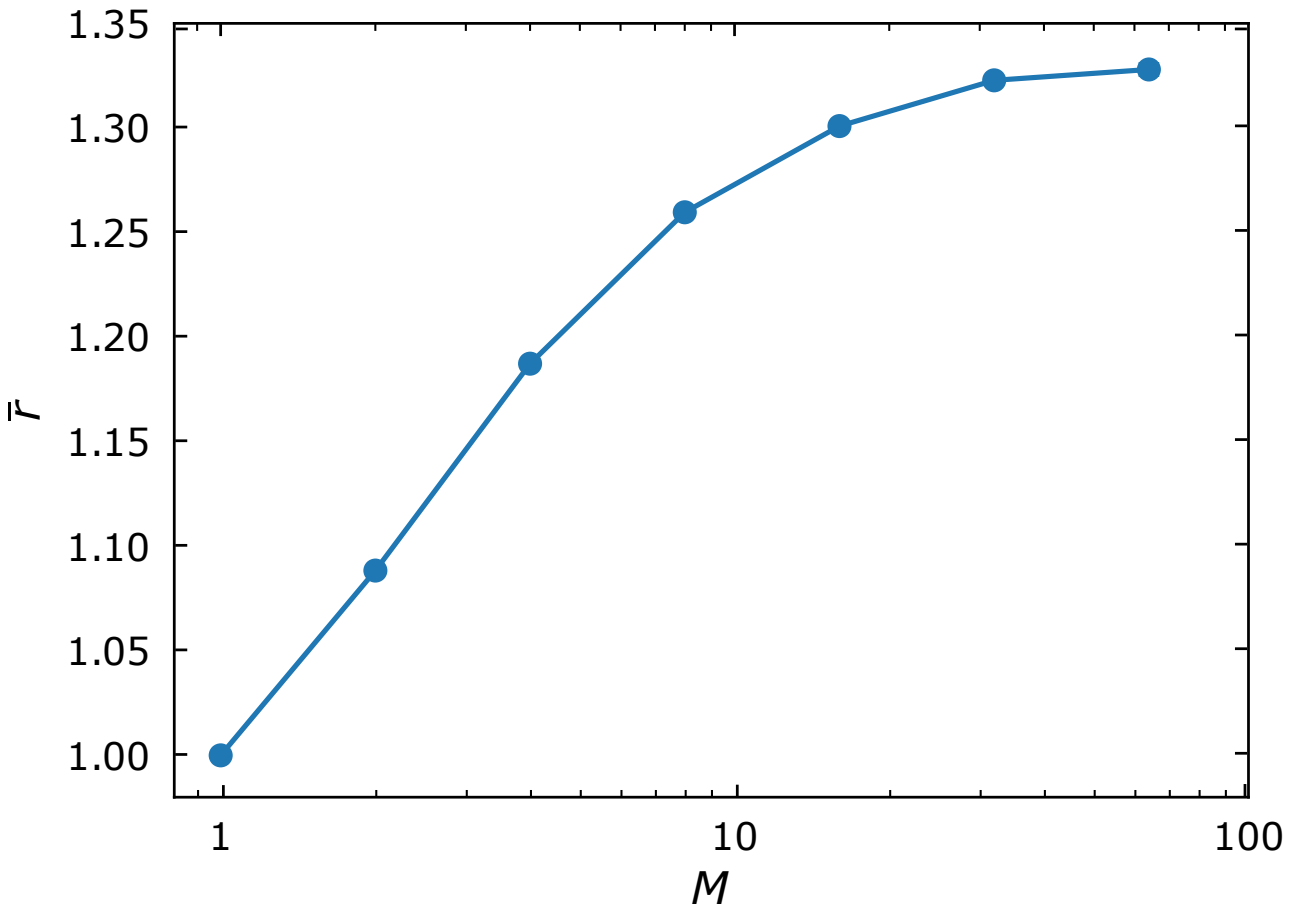

**Fig. 5.** Estimate of the effective influence of the noise correlations for the E noise images for test Ic and S/N ≃ 30. The $\bar{r}$ factor is plotted as a function of the side length $M$ of the sub-regions of 2 000 noise images and the error bars are the standard error of the mean. In this case, the value of $\bar{r} = 1.32$ for large values of $M$ gives the factor by which the S/N is overestimated when ignoring the noise correlations.

In order to estimate the effective influence of the noise correlations on the test environment, in particular for the E images, we use the approach of Hartlap et al. (2009). We create 2000 pure noise images (no galaxy within them) of a size of $512 \times 512$ pixels, which are processed in the same manner as the galaxy images (see Sect. 3). We add different amounts of extra Gaussian noise to the images, with dispersions $\sigma_{\rm G}$ [ADU] ranging from 1.0 to 6.0 with a step of 0.2, and we run the testing environment for each of these values. For each run, we estimate the $\bar{r}$ factor defined as

$$\bar{r}\,(M) = \left\langle \frac{\sigma^{\rm measure}_{N,i}}{M^2 \sigma^{\rm measure}_{1,I}} \right\rangle_i , \tag{7}$$

where $\sigma^{\rm measure}_{N,i}$ is the SD of the pixel sum measured in independent quadratic sub-regions within noise image $i$ with side length $M = \sqrt{N} \in \{1, 2, 4, 8, 16, 32, 64\}$ pixels, and $\sigma^{\rm measure}_{1,i}$ is measured on single pixel values, averaging over all the pure noise fields. In the absence of correlated noise, the $\bar{r}$ factor would be equal to 1 for all $N$. However, in the presence of noise correlations and for large $N$, the $\bar{r}$ factor converges to the value by which $\sigma^{\rm measure}_1$ under-estimates the uncorrelated dispersion in a single pixel. Figure 5 shows an example of the measured $\bar{r}\,(M)$ for test I. In this case, the ordinary noise measure based on the single pixel dispersion, which ignores the noise correlation, will overestimate the S/N of the E galaxies by a factor of $\bar{r} = 1.32$. We note that this convergence is expected since the size of the quadratic regions considered becomes much larger for large $N = M^2$ than the scale on which the noise correlation occurs. This results in asymptotically uncorrelated estimates of the dispersions, $\sigma^{\rm measure}_N$, which represents the asymptotic value of $\sigma^{\rm measured}_{N,i}$.

The bias of a shear measurement algorithm is sensitive to the S/N of a source. To make measurements on the E and D images comparable, while acknowledging that the E images contain correlated noise, we tried tuning the Euclidisation procedure so that the galaxies share the same true S/N in E and D. Given the correlated noise, this is not achieved when E and D have the same flux and the same $\sigma_1$. Instead, we adjusted the amount of additional white noise $\sigma_{\rm G}$ added to E so as to match $\sigma_{\rm N}$ between E and D. This is achieved when the ratio $\sigma_{\rm 1D}/\sigma_{\rm 1E}$ reaches the value of $\bar{r}$ obtained from Eq. (7). In practice, we determine $\sigma_{\rm G}$ so as to obtain $\bar{r} = \widehat{r}$, as illustrated in Fig. 6, with

$$\widehat{r} = \frac{\sigma_{\rm 1D}}{\sigma_{\rm 1E}}\,, \tag{8}$$

where $\sigma_{\rm 1D}$ and $\sigma_{\rm 1E}$ are the single-pixel dispersions in the images. The intersection between the two curves, representing fits to the estimates defined in Eqs. (7) and (8), will provide the best estimate for the $r$ factor, for instance, $r^{\rm true}$, along with the $\sigma_{\rm G}$ value that we will use in the noise correlation correction of the testing environment. Figure 6 provides a visual example of this intersection, indicating how we estimate $r^{\rm true}$ for test I.

[10] We note that despite using deeper HST data, we would continue to limit galaxies to the same criterion, specifically the $I_{\rm E} < 24.5$ mag selection, without including additional fainter sources.

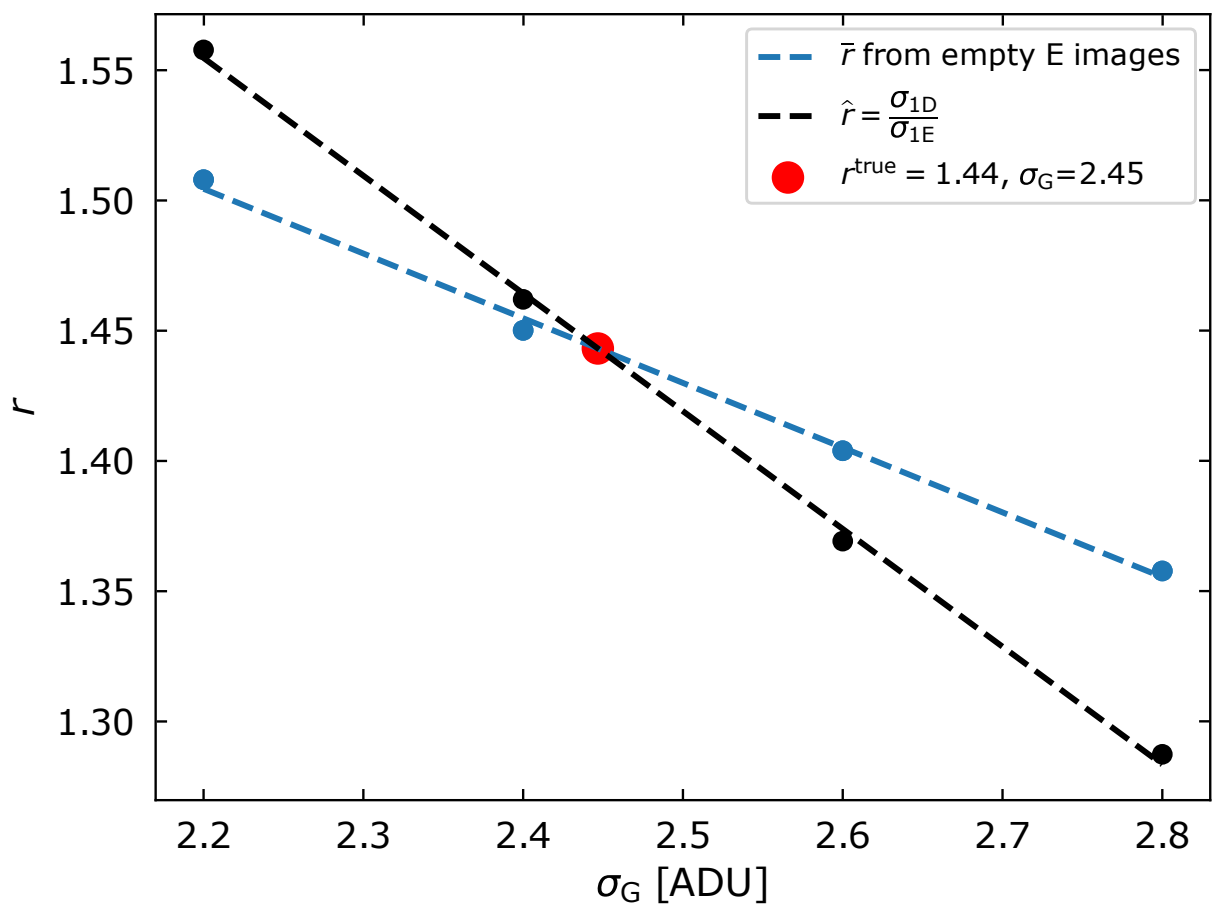

**Fig. 6.** Estimate of the $r$ factor corresponding to the actual value of $\sigma_{\rm G}$ we need to use in the testing environment to both account for the noise correlation and match the noise properties of the E images with the D images. The blue and the black dotted curves show the linear fits to the quantities defined in Eqs. (7) and (8), which were then used in test Ic. Their intersection defines the estimated value for $r^{\rm true}$ and the corresponding $\sigma_{\rm G}$ (red dot).

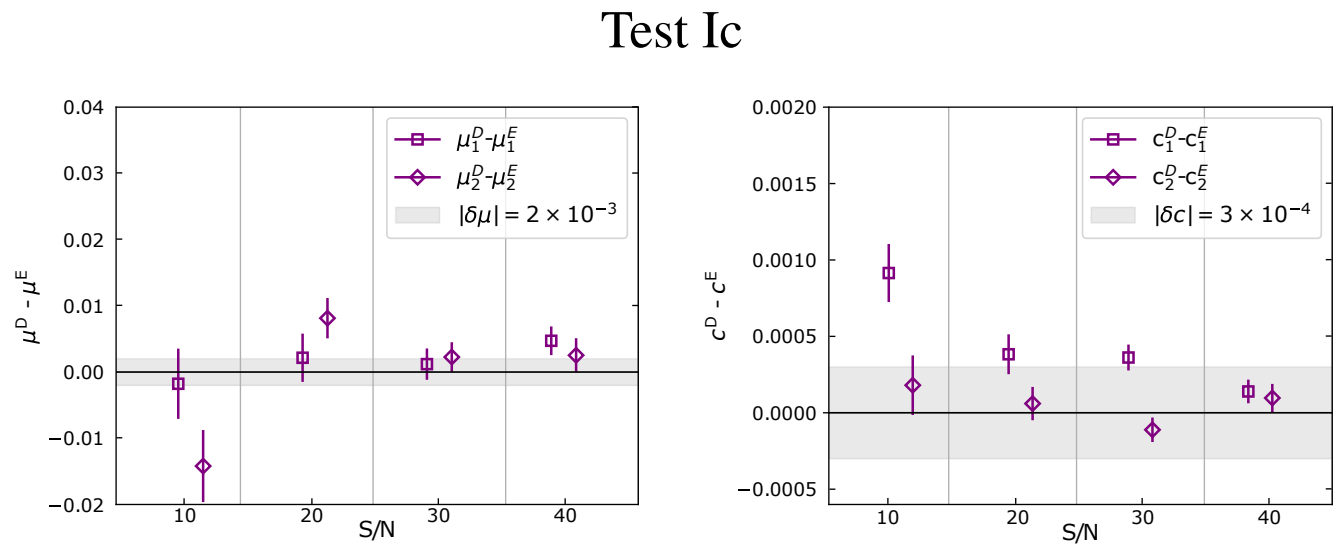

**Fig. 7.** Correction for the noise correlation and interpolation kernel. Multiplicative (left panels) and additive (right panels) shear bias differences obtained employing a *Euclid*/VIS pixel scale of 0.″1, a `lanczos15` interpolation kernel, when we account for the impact of noise correlations in the S/N matching between the D and E images (test Ic), as described in Sect. 5.2. The data points show the bias obtained from the fit with the error bars indicating $1\,\sigma$ uncertainties for a number of galaxies $N_{\rm gal} \simeq 1.4\times10^6$ for each S/N. For comparison, the shaded area indicates the requirements on uncertainty on the $\mu$ and $c$ biases arising from the total shear measurement process for an *Euclid*-like survey (Cropper et al. 2013).

The results for test I, which have been adjusted to account for the influence of correlated noise, are depicted in Fig. 7 (hereinafter referred to as 'test Ic', with 'c' indicating the noise correction). Furthermore, in accordance with the approach outlined by Kannawadi et al. (2021), for this test, we opted to employ the `lanczosN` interpolation scheme with the parameter $N$ set to 15, as opposed to the default `quintic` option provided by `GalSim`. The rationale behind this choice is that the `lanczos15` scheme enhances accuracy without significantly slowing down the process of Euclidisation. We find that `lanczosN` with `N` = 15 strikes an optimal balance between speed and precision. In Fig. 7 we observe a reduction in the multiplicative bias difference to less than 1%. For reference, the shaded area represents the requirements on uncertainty on the $\mu$ and $c$ biases arising from the total shear measurement process for an *Euclid*-like survey (Cropper et al. 2013). This reduction in bias can be ascribed to two factors: the noise correction and the Lanczos15-kernel, which possess a support region significantly smaller than the usual postage stamp sizes we encounter. In addition, with this attempt to mitigate the effect of the correlated noise, the trend for $\Delta\mu$ becomes flatter at lower S/N. We also observe a minor reduction in the difference in additive bias. We identify a factor that likely contributes to the observed non-zero value of $|\Delta\mu|$. It is linked to the sensitivity of $|\Delta\mu|$ to the $\sigma_{\rm G}$ value, which will be discussed further in Sect. 5.3. We used a fit of the galaxy model to estimate the S/N of the D and $\rm E_{uncorrected}$ images. For the E images, we also derived a 'corrected' S/N (labelled as $\rm E_{corrected}$ in Fig. 9), which takes into account the correlation of the pixel noise, by scaling the noise by an $r$ factor, as shown in Eq. (8). Figure 9 shows the comparison of the mean values over $N_{\rm gal} \simeq 10^4$ galaxies for the measured $\rm S/N_{fit}$ (which is biased in the presence of correlated noise) as a function of the input half-light radius $R_{\rm e}$. As expected, the $\rm S/N_{fit}$ for the $\rm E_{corrected}$ images is lower after accounting for the noise correction compared to the S/N value in the $\rm E_{uncorrected}$ images prior to the correction and it is compatible with the $S/N_{\rm fit}$ for D.

For this test, we also inspected the multiplicative and additive bias as a function of $R_{\rm e}$ and $n$ for the highest S/N case, representative of other S/N levels. Figure 8 (left panel) shows $\mu_1$ and $\mu_2$ across the bins, indicating that the biases generally remain close to zero, with deviations between the two multiplicative bias components within error bars, suggesting that the measurements are consistent across a range of galaxy properties. However, there are a few bins where the deviations increase slightly, which could indicate areas requiring further investigation. The bottom panel of Fig. 8 presents the additive bias components, as a function of the same properties. The additive bias also remains around zero, consistent with no significant systematic offset. The scatter and error bars suggest the presence of minor fluctuations, but overall, there is no clear trend indicating a significant dependence on $R_{\rm e}$ or $n$.

### 5.3. Sensitivity to $\sigma_{\rm G}$

The uncertainty in determining $\sigma_{\rm G}$ propagates into the uncertainty of $\Delta\mu$. To estimate this dependency, we test the influence of uncertainties in $\sigma_{\rm G}$ on the differences in multiplicative and additive biases by varying its value from 2.5 to 2.7 (i.e. of about 10%) for test Ic. These are denoted as 'test Ic with $\sigma_{\rm G}$ = 2.5' and 'test Ic with $\sigma_{\rm G}$ = 2.7'. In these cases, we employed the default `quintic` interpolation kernel.

As shown in Fig. 10, this change in $\sigma_{\rm G}$ can lead to a percent-level difference in $\Delta\mu$. From Fig. 6, we can estimate the actual uncertainty on our determination of $\sigma_{\rm G}$ to be on the order of a few percent. We conclude that the uncertainty on $\Delta\mu$ from this determination of $\sigma_G$ alone is of the order of a few tenths of a percent. Given these promising results, it will be necessary to carefully investigate the inference of $\sigma_G$ (as emphasised in Sect. 5.2) in the case where such an approach is taken to mitigate the effect of noise correlation in the Euclidisation.

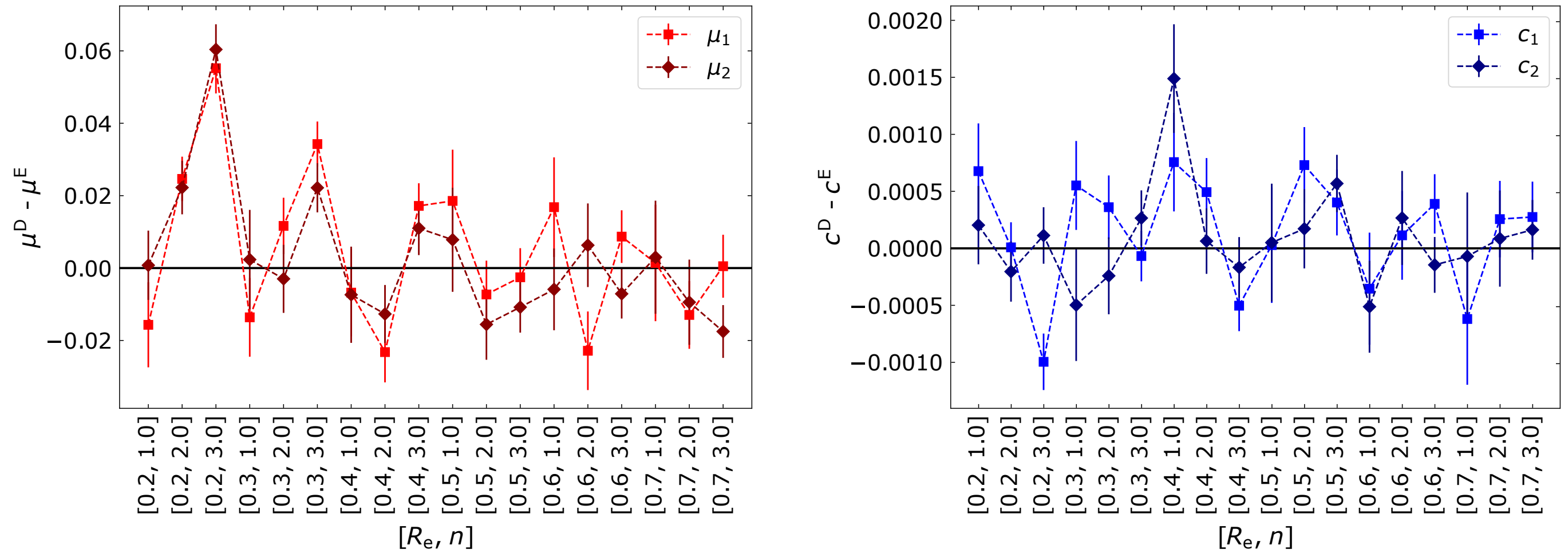

**Fig. 8.** Multiplicative (left panel) and additive (right panel) shear bias differences as functions of effective radius ($R_e$) and Sérsic index ($n$), using a *Euclid*/VIS pixel scale of 0.″1 and a `lanczos15` interpolation kernel. The analysis accounts for noise correlations in S/N matching between the D and E images (test Ic), as detailed in Sect. 5.2. The data points show the bias obtained from the fit with the error bars indicating 1 $\sigma$ uncertainties for a number of galaxies $N_{gal}$ for S/N $\simeq$ 40.

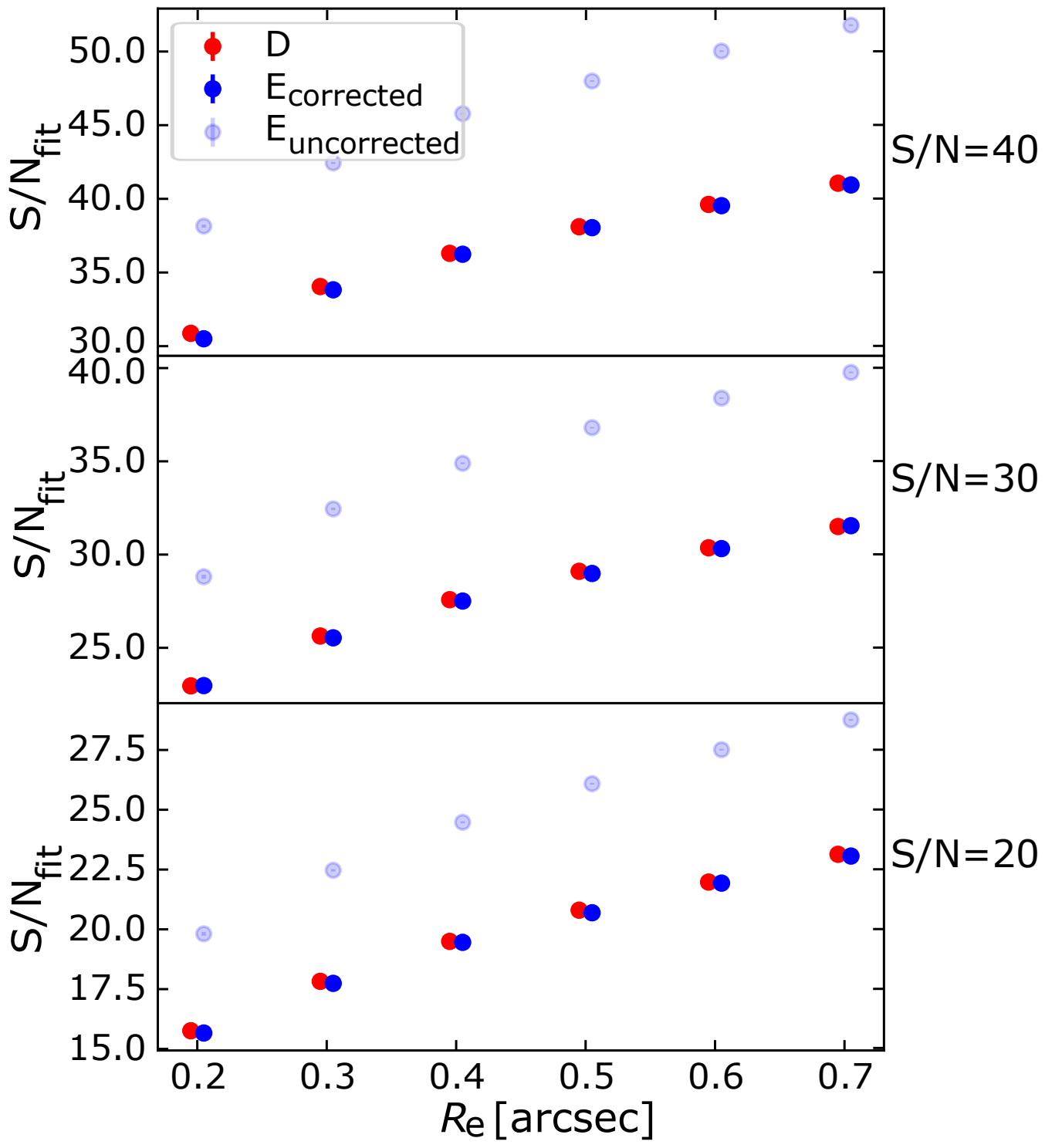

**Fig. 9.** Results from the galaxy model fits using the native HST/ACS and *Euclid*/VIS pixel scales, for a number of galaxies $N_{gal} \simeq 10^4$, as described in Sect. 5.2 (test Ic). The fitted parameter $S/N_{fit}$ (which is biased in the presence of correlated noise) is shown as a function of the input $R_e$ for both the output galaxies D, $E_{uncorrected}$, and for the $E_{corrected}$ images accounting for the noise correlation, for three values of the S/N. The data points are the average over the number of galaxies in each sample. The corresponding $1\sigma$ uncertainties are smaller than the size of the points.

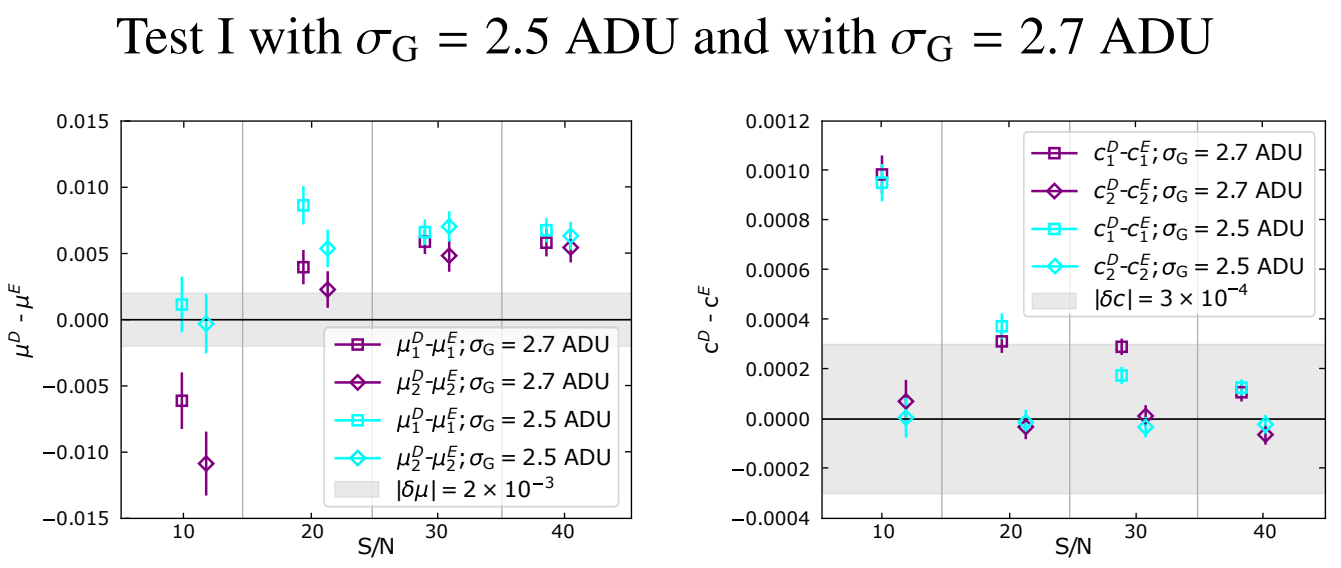

**Fig. 10.** Sensitivity test to $\sigma_G$. Multiplicative (left panel) and additive (right panel) shear bias differences obtained employing a *Euclid*/VIS pixel scale of 0.″1 and a `quintic` interpolation kernel, as described in Sect. 5.3, when we use extra Gaussian noise with $\sigma_G$ = 2.5 ADU (test I with $\sigma_G$ = 2.5 ADU), and a $\sigma_G$ = 2.7 ADU (test I with $\sigma_G$ = 2.7 ADU). The data points show the bias obtained from the fit with the error bars indicating 1 $\sigma$ uncertainties for a number of galaxies $N_{gal} \simeq 8.9 \times 10^6$ for each S/N value. For comparison, the shaded area indicates the requirements on uncertainty on the $\mu$ and $c$ biases arising from the total shear measurement process for an *Euclid*-like survey (Cropper et al. 2013).

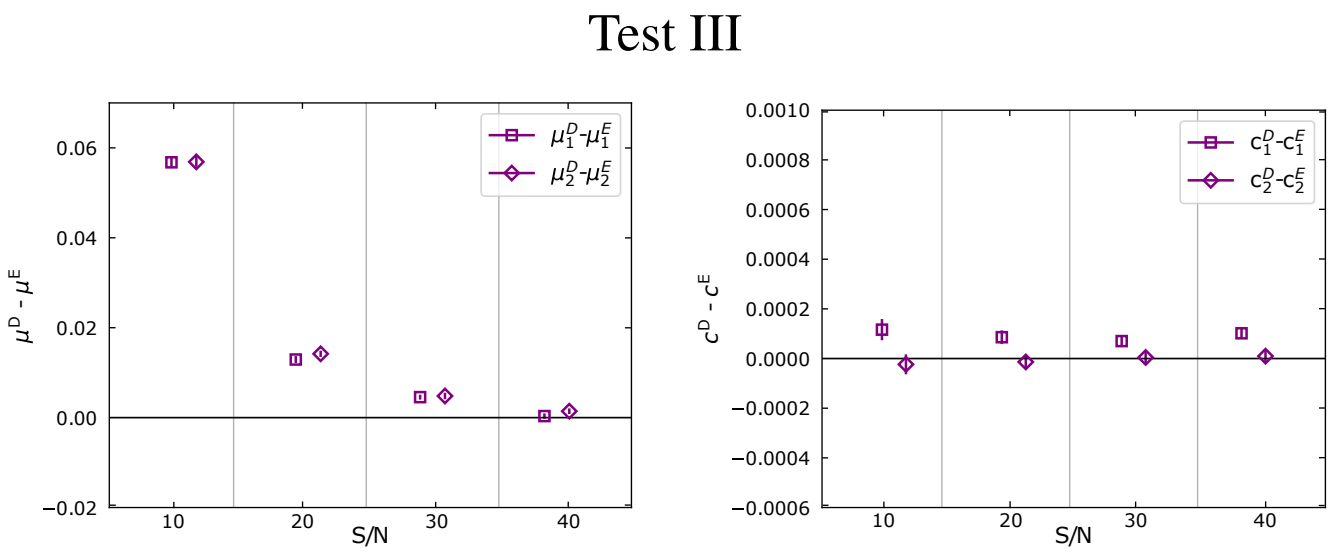

**Fig. 11.** Effect of pixel sampling. Multiplicative (left panel) and additive (right panel) shear bias differences obtained when deliberately using a pixel scale of 0.″04, as described in Sect. 5.4 (test III). The data points show the bias obtained from the fit with the error bars indicating 1 $\sigma$ uncertainties for a number of galaxies $N_{gal} \simeq 33.8 \times 10^6$ for each S/N.

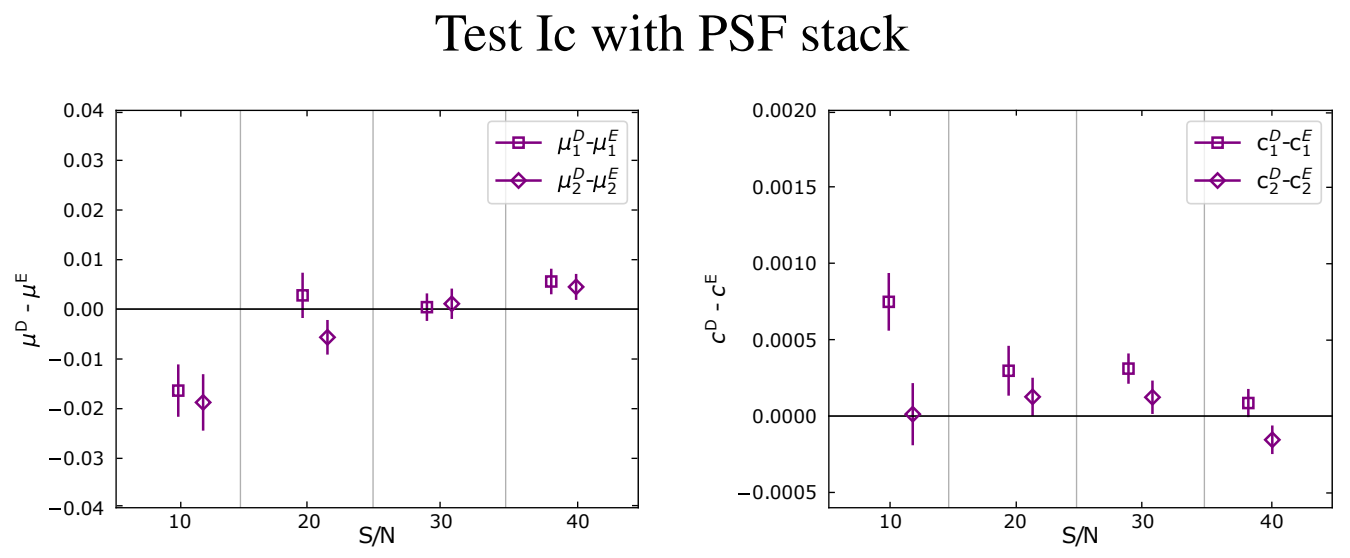

**Fig. 12.** Impact of PSF model errors. Multiplicative (left panel) and additive (right panel) shear bias differences obtained employing a *Euclid*/VIS pixel scale of 0.″1 (test Ic with PSF stack). For the analysis shown here we have employed a star stack as PSF model for the convolution and a model stack for the deconvolution, as described in Sect. 5.5. The data points show the bias obtained from the fit with the error bars indicating $1\,\sigma$ uncertainties employing $N_{\rm gal} \simeq 18.3 \times 10^6$ galaxies for each S/N value.

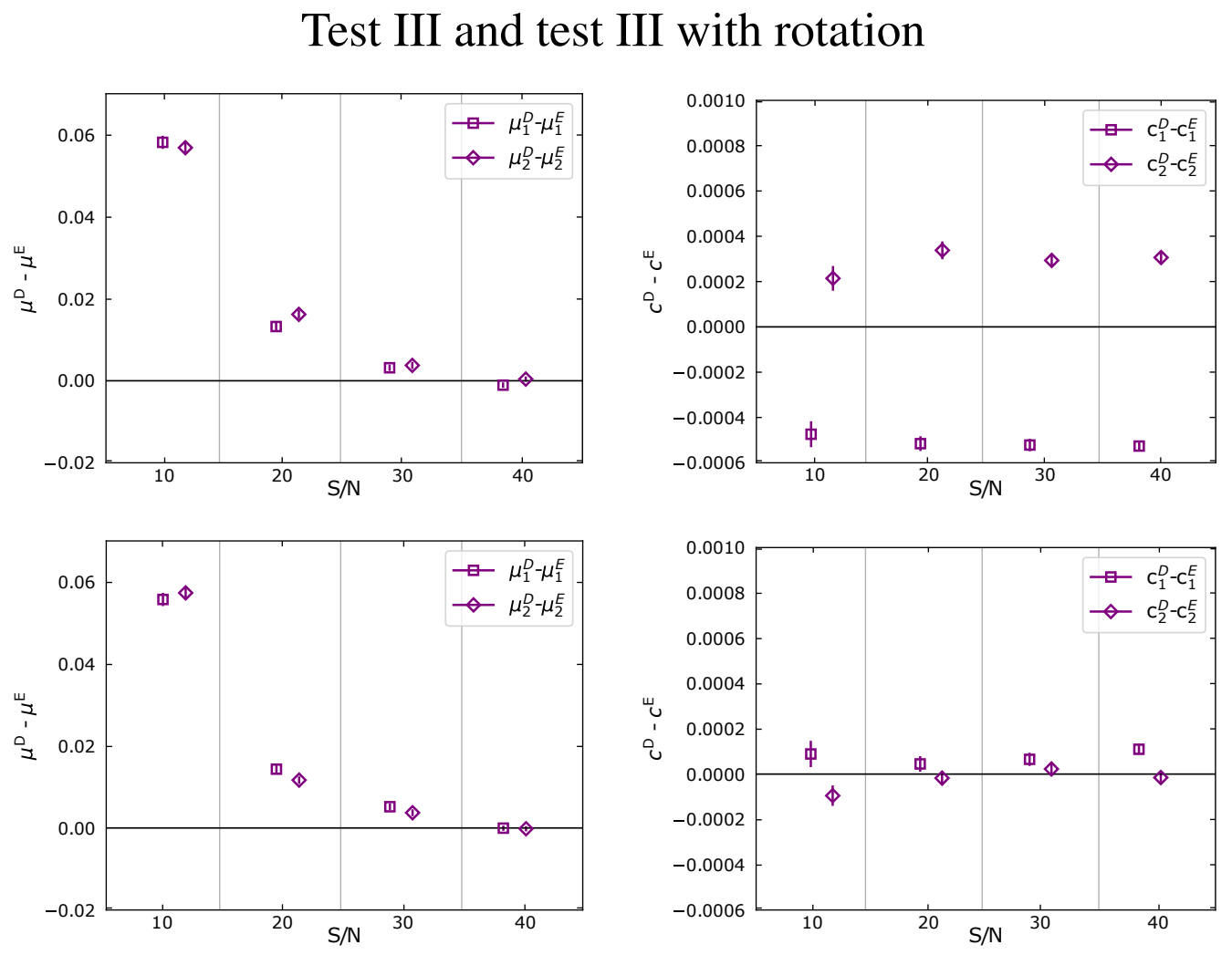

**Fig. 13.** Effect of HST PSF error and post-deconvolution isotropisation. Multiplicative (left panel) and additive (right panel) shear bias differences obtained when using two different HST `TinyTim` PSF models for HST conv and HST deconv and a pixel scale of 0.″04 (upper panels, test III), as described in Sect. 5.5. The lower panels show the results applying an extra rotation after the deconvolution (test III with rotat.). The data points show the bias obtained from the fit with the error bars indicating $1\,\sigma$ uncertainties for a number of galaxies $N_{\rm gal} \simeq 24.7 \times 10^6$ for each S/N value.

### 5.4. Use of a finer pixel scale

In order to analyse the impact of sampling on the Euclidisation, we perform an experiment referred to as 'test II', where we deliberately set both the HST and the *Euclid*/VIS pixel scale to 0.″04, instead of the native values. This value is close to the native HST/ACS pixel scale but a bit finer (e.g. matching the sampling of HST/UVIS). We therefore need to over-sample the output galaxies by a factor of 2 prior to running KSB, to match the sampling of the PSF, and also convolve this PSF by a 0.″04-wide top-hat pixel profile. For this test we employ the default `GalSim` `quintic` interpolation kernel.

As shown in Fig. 11 (left panel), we find that the finer sampling strongly reduces the $\mu$ bias at higher VIS S/N ($\geq$ 20), reaching $\Delta\mu$ of the order of $10^{-3}$ at S/N $\simeq$ 40. This is an important result of our analysis since the multiplicative bias difference converges to zero at high S/N, where noise-related biases are expected to be small. However, we suspect that a higher noise value at S/N $\simeq$ 10 is also responsible for the increase in the multiplicative bias difference. Thus, the combination of finer sampling and deeper HST images could be an alternative solution (see Sect. 5.2) to recover the accuracy on the $\mu$ bias we desire. While for the $c$ bias difference, 11 (right panel), there is no impact and they are in agreement with what we have found in the previous test. We also perform the galaxy model fits on samples of around $10^4$ galaxies for each S/N. We find a good agreement in size and flux between the D and E galaxies. However, for $n_{\rm fit}$ the fitted values reveal discrepancies comparable to those of Fig. 3 see Sect. 5.1.

When using the finer pixel scale, the compensation for noise correlation as described in Sect. 5.2 cannot be applied, as the pixel noise after its isotropisation is already higher than the noise in the direct branch.

### 5.5. Different HST PSF models

In order to evaluate the impact of the HST PSF model uncertainties on the use of HST images, we analyse a set-up where we employ moderately different HST PSF models for the convolution and the deconvolution in the bottom branch of Fig. 1. We probe the sensitivity to PSF uncertainties in two set-ups.

In the first set-up, to which we refer to as 'test Ic with PSF stack', for the convolution and the deconvolution (see the bottom of Fig. 1), we used an average star stack and an average model stack, respectively, as PSF models. We obtained these from the pipeline presented in Gillis et al. (2020) that we adapted for our scope in Sect. 6.1. In this set-up, we employed the native *Euclid*/VIS and HST/ACS pixel scales, the `lanczos15` interpolation kernel. We then implemented the noise correlation correction.

Figure 12 (compare to Fig. 7) shows that the multiplicative bias difference is in agreement with what we found in test Ic (see details in Sect. 5.2), although the $\Delta\mu$ becomes negative at S/N = 10. The residuals between the model and the star stacks (see Sect. 6.1) seem to affect the galaxy measurements especially at lower S/N. The additive bias difference is consistent with Fig. 7.

In the second set-up, which we refer to as 'test III', we considered two `TinyTim` PSF models in filter F814W, which we created at different centre positions $(x, y) = (1088, 488)$ and (64, 64), and foci of 3.1 and 1.5 $\mu$m. The difference between these models is at a level similar to typical systematic uncertainties of the PSF model (see Sect. 6). Therefore, their use allows us to gauge the approximate level of the impact of systematic HST PSF model errors on the Euclidisation set-up.

In this test, we employed a finer pixel scale of 0.″04 for each step of the procedure and the *Euclid* PSF being directly passed to the KSB method with a pixel scale of 0.″04. In addition, we employed the `quintic` interpolation kernel. The results now show a substantial $c$ bias difference between D and E (see the top right panel of Fig. 13). The $\mu$ bias difference behaves similarly as in Fig. 11 (see Sect. 5.4). This test suggests that typical ACS PSF model uncertainties have little impact on the $\mu$ bias calibration, but could significantly affect the $c$ bias calibration, in agreement with Semboloni et al. (2013).

To avoid this $c$ bias issue, we propose a 'post-deconvolution isotropisation' (PDI) of the HST images, 'test III with rotat.'. It consists of adding an extra random rotation within the Euclidisation procedure, after the deconvolution by the HST PSF and prior to the application of the shear. This rotation helps cancel the additive bias induced by the anisotropy of PSF. In order to keep the shape noise cancellation, the *Input galaxy* pair must be rotated by the same random rotation angle, drawn from a uniform distribution of values between 0 and 180 degrees. As shown in the bottom panels of Fig. 13, this indeed sufficiently suppresses the $c$ bias difference, thereby resolving the issue. It is worth bearing in mind that, when we use real HST data, the input galaxies for the Euclidisation set-up correspond to the HST conv images of Fig. 1. In this case, the PDI is included not only to decorrelate the analysis from the HST PSF anisotropy residuals, but also because we have a finite number of HST galaxies. Indeed, for each galaxy we want to be able to generate output galaxies with all kinds of rotations as is usually done in WL image simulations (e.g. Mandelbaum 2018).

Furthermore, as the influence of sampling might be related to the PSF size, which can smooth the galaxy differently, we repeat the experiment using a different and slightly narrower HST PSF from the F606W filter instead of from the F814W filter. In this case, both the convolution and the deconvolution in the bottom branch of Fig. 1 are performed with the same F606W filter PSF. We find that the use of different filters for the HST PSF does not affects the results, with both components of $\Delta\mu$ and $\Delta c$ are consistent with each other at each $S/N$.

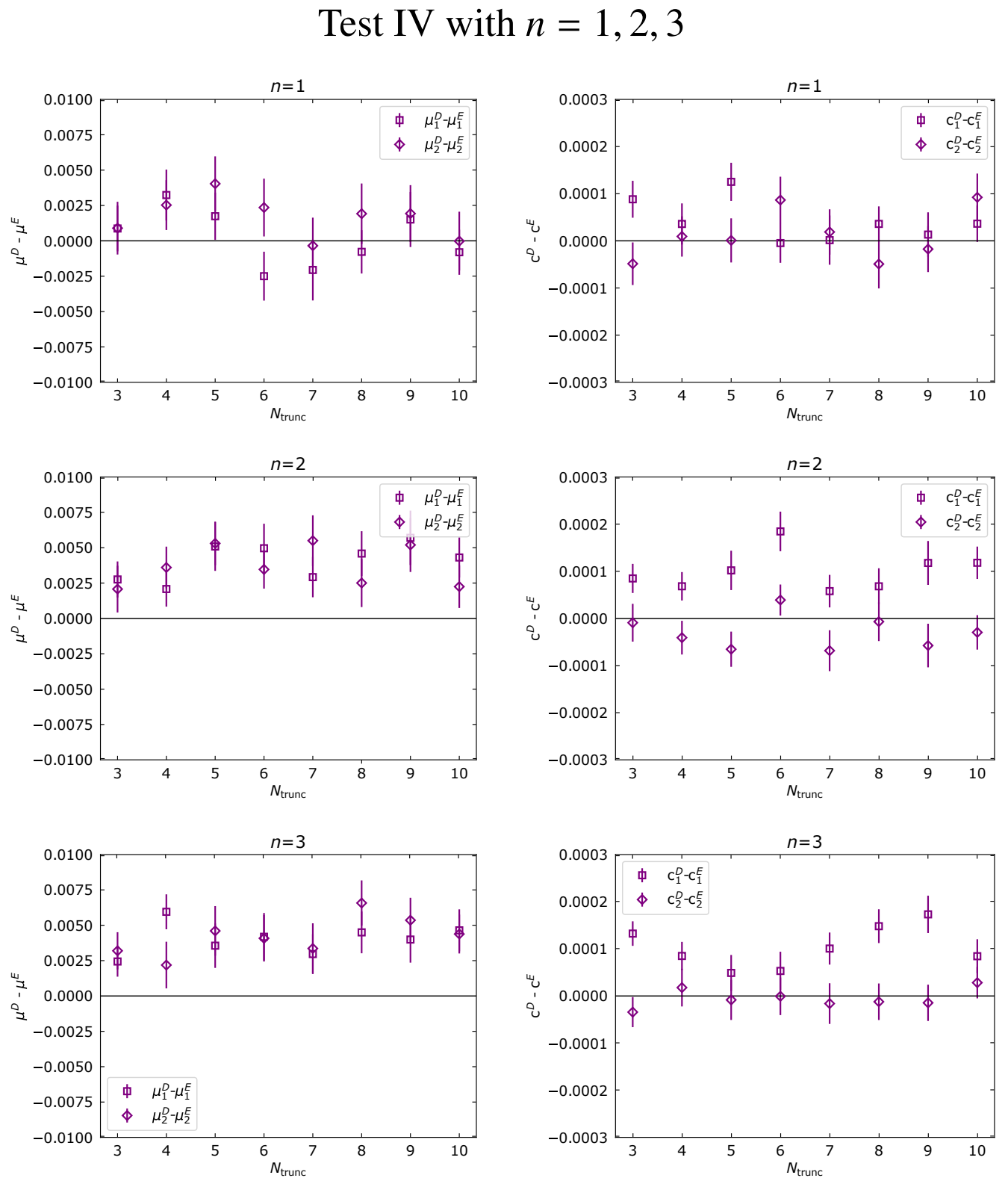

**Fig. 14.** Analysis of the impact of introducing a truncation radius for input galaxies. Multiplicative (left panel) and additive (right panel) shear bias differences obtained when using a truncation radius for the input galaxies, a pixel scale of 0.″04, and a `quintic` interpolation kernel (test IV with $n = 1, 2, 3$), as described in Sect. 5.6. In each row, the results for a specified Sérsic index $n$ are shown. The data points show the bias obtained from the fit with the error bars indicating 1 $\sigma$ uncertainties for a number of galaxies $N_{\rm gal} \simeq 77.0 \times 10^6$ for S/N $\simeq$ 30.

### 5.6. Truncation radius for the input galaxies

Background noise makes the faint outer parts of galaxy brightness profiles undetectable. As shown by Hoekstra et al. (2021), shape measurement biases may depend on the external regions of a galaxy. For example, a potential outer truncation of the brightness profiles would affect shape calibrations at a level relevant for experiments such as *Euclid*. Given the presence of noise, it is difficult to quantify such a potential outer truncation radius accurately. This introduces systematic uncertainties in shear calibrations that use model galaxies described by analytic brightness profiles, or that rely on simulated galaxy images in some way.

In this subsection, we investigate whether calibration simulations based on HST postage stamps (rather than analytic galaxy models) can help to avoid this issue. For this, we employed the testing environment, using simulated input galaxies that have different truncation indices, defined as $N_{\rm trunc} = R_{\rm trunc}/R_{\rm e}$, which we varied in the range [3, 10] in one unit increments. For 'test IV ($n \in \{1, 2, 3\}$)', as we labelled it, we restricted our analysis to $R_{\rm e}[''] \in \{0.3, 0.4, 0.5, 0.6, 0.7\}$ and $|g_i| < 0.04$, with $i = 1, 2$. The whole testing environment uses a pixel scale of 0.″04 and the default `GalSim` interpolation scheme.

Figure 14 illustrates that multiplicative bias differences are indeed independent of $N_{\rm trunc}$. Thus, the Euclidisation set-up yields an accurate multiplicative bias calibration independent of what the true galaxy truncation radius may be. In the same figure (Fig. 14), we also see a disagreement between the additive bias components for some $N_{\rm trunc}$, which is worth being further investigated. Nevertheless, no clear trend or significant dependence on $N_{\rm trunc}$ is detected overall.

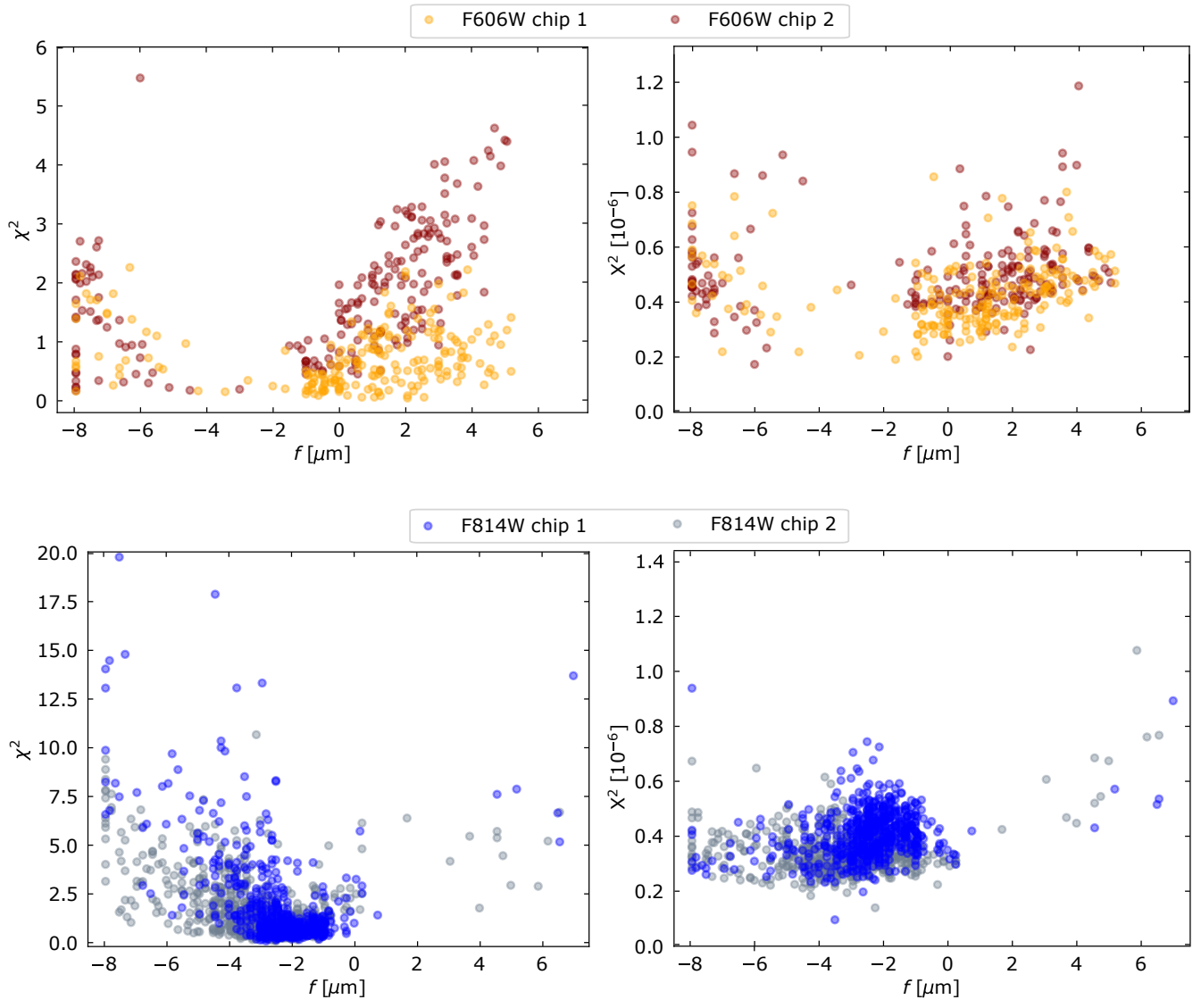

**Fig. 15.** Statistical parameter $\chi^2$ (left panels) and the quality-of-fit parameter X$^2$ (right panels) for each filter and chip plotted against the best-fit focus value of each of 205 star fields in the F606W filter and 645 star fields in the F814W filter.

## 6. Analysis of the accuracy of the `TinyTim` PSF model for the *Hubble* Space Telescope

Accurate estimation of WL shear relies on correcting observational and instrumental effects to ensure unbiased measurements of galaxy ellipticities. These corrections encompass various factors, including the convolution of images with the PSF of the telescope, geometric distortion, particularly significant in the ACS camera due to its off-axis position on the HST, and charge transfer inefficiency (CTI) during CCD readout (Massey 2010). Additionally, temporal variations in the ACS PSF, caused by thermal fluctuations during the telescope's orbit, further complicate correction efforts (Lallo et al. 2006; Rhodes et al. 2007; Schrabback et al. 2007). Ensuring accurate corrections necessitates reliable PSF models. Building on previous research (Rhodes et al. 2007; Gillis et al. 2020), we investigate the accuracy of HST/ACS PSF models generated using `TinyTim` (Krist et al. 2011). These models are crucial not only for direct WL measurements based on HST observations but also for WL image simulations utilising HST galaxy observations as input. Deconvolution of observed galaxy shapes from the PSF in the Euclidisation process requires precise knowledge of the PSF shape and its evolution over time, given the cyclical expansion and contraction of the HST telescope barrel during orbit (Lallo et al. 2006; Rhodes et al. 2007). These fluctuations, due to the telescope's slow breathing, alter the distance between primary and secondary mirrors, affecting focus and subsequently, the PSF size and shape. As a result, the evolving PSF characteristics impact observed galaxy ellipticities.

Generally speaking, errors in the size of the PSF model introduce both multiplicative and additive biases, while errors in the PSF model shape and anisotropy also introduce additive biases (Paulin-Henriksson et al. 2008). In Sect. 6.1, we focus on generating and investigating the accuracy of `TinyTim` PSF models for the HST/ACS. This analysis involves dense stellar fields in the F814W filter, as well as the F606W filter images, which were already investigated in Gillis et al. (2020). The aim is to quantify the impact on the testing environment results. Additional analyses on the `TinyTim` PSF models in the regime of low stellar densities, as in the galaxy fields, are reported in Sect. 6.2. In addition, in Appendix B.2 we compare `TinyTim`-based estimates of the HST telescope focus in stellar fields to the first coefficient in the Principal Component Analysis (PCA) of ACS PSF variations by Schrabback et al. (2010; 2018).

### *6.1. Generation and analysis of TinyTim PSF models*

To generate HST/ACS PSF models, we used `TinyTim` (Krist et al. 2011), a standard tool for generating PSF models for the HST. Gillis et al. (2020) conducted an in-depth analysis of the accuracy of `TinyTim` PSF models for HST/ACS images taken in the F606W filter by comparing them to stellar images in observations of star fields. While accounting for the PSF dependence on position and telescope focus, this analysis revealed significant residual differences between the models and stars. They therefore computed an updated set of HST PSF Zernike coefficients from the star field observations, which allowed them to reduce, but not completely remove, these residuals. Along the same line, we extend the wavelength analysis to images in filter F814W.

The methodology of Gillis et al. (2020) is computationally expensive, as it creates an individual `TinyTim` PSF model for every star of the analysis. In the present paper, we chose instead to pre-compute `TinyTim` models on a grid of positions and focus parameters and query this database for the nearest pre-computed neighbor to every observed star. We subdivided the images of each chip ($4096 \times 2048$ pixels) into cells of $128 \times 128$ pixels in size and computed them in the focus range $-10\,\mu$m to 8.5 $\mu$m (exceeding the range of expected variations, see Gillis et al. 2020) in steps of 0.1 $\mu$m. `TinyTim` generates finely over-sampled models with a subsampling factor of 8, not accounting for the convolution with a charge diffusion kernel. This factor is large enough to allow us to shift the PSF model to optimally[11] match the proper subpixel centre for any star.

Since the `TinyTim` PSF models using the default parameters fail to adequately characterise the observed PSF, we used the refined best-fit estimates of higher-order Zernike coefficients from Gillis et al. (2020). These coefficients characterise the optics of the telescope, for instance, the focus offset corresponds to the fourth Zernike polynomial's coefficient. For this analysis, we fit our gridded pre-computed PSF models to stars in a set of HST/ACS star fields described in Schrabback et al. (2018), which comprise 205 star fields exposures observed in F606W and 645 star fields exposures observed in F814W. Compared to the computation of a PSF model for each individual star, the computation of PSF models on a grid yields a substantial speed-up, while yielding consistent focus estimates (with a difference of the order of $10^{-4}\mu$m). Each observed star is compared to the models from the corresponding cell, in terms of a quality-of-fit statistic, $\mathrm{X}^2$, of the fitting residuals (see Appendix B.1). The $\mathrm{X}^2$ is a chi-squared-like combination of a heterogeneous set of statistics, weighted in accordance with their expected impact on shear estimation bias, with a lower $\mathrm{X}^2$ corresponding to less bias in shear estimates. This statistic is based on high-order moments of the PSF fitting residuals, which explicitly summarises the effect of PSF model mismatch on shear measurements (for details see Gillis et al. 2020). Thus, the best-fit focus corresponds to the value which minimises the $\mathrm{X}^2$. In addition, we also computed a $\chi^2$ value based on the quadrupole moments of the brightness distribution of the PSF, as detailed in Gillis et al. (2020).

We show the resulting $\chi^2$ and $\mathrm{X}^2$ for all star fields as a function of the best-fit focus in Fig. 15. Overall, the range of $\chi^2$ and $\mathrm{X}^2$ is similar in both filters, suggesting that the refined Zernike coefficients from Gillis et al. (2020) perform similarly well for F606W (on which they were calibrated) and F814W. However, we note that the F606W observations show a broader scatter in the recovered best-fit focus. We suspect that this mostly reflects a different range in typical observing conditions of the F814W and F606W star fields. A similar behavior was also observed in Gillis et al. (2020). They conducted tests on PSF fitting using simulated fields under various conditions. Their findings indicate that among the effects tested, only the addition of unresolved binaries has a noticeable impact. Specifically, it biases the fitted focus value by up to about $1\mu$m away from the best-fit focus, which is close to $-3\mu$m. This occurs because binaries are randomly oriented, resulting in these stars appearing larger collectively. The PSF fitting method cannot fully accommodate this phenomenon, but it must adjust somehow, typically by modifying the parameter that best approximates this effect. Consequently, a biased focus offset value is fitted, and adjusting this value does not precisely account for the presence of binary stars. Including the influence of binary stars in the fitting procedure will be pursued in future investigations. Furthermore, Gillis et al. (2020) assessed the impact that size errors are likely to have on shear bias measurements, and found that they are more than an order of magnitude below the level which would cause issues for the *Euclid* mission (see their Fig. 7, which shows the fit statistics for their $Q_\mathrm{S}$ parameter, a proxy for size, in the centre-right and bottom-right panels). We thus conclude that this factor is unlikely to be of concern here. For illustration, in Fig. 16 we show the stacks of the observed stars, of the model PSF for a star field in the

[11] A greater sub-sampling factor was found to have negligible benefit, see Gillis et al. (2020).

F606W and F814W filters. For both filters, the residuals are at a moderate level, but detected with high significance.

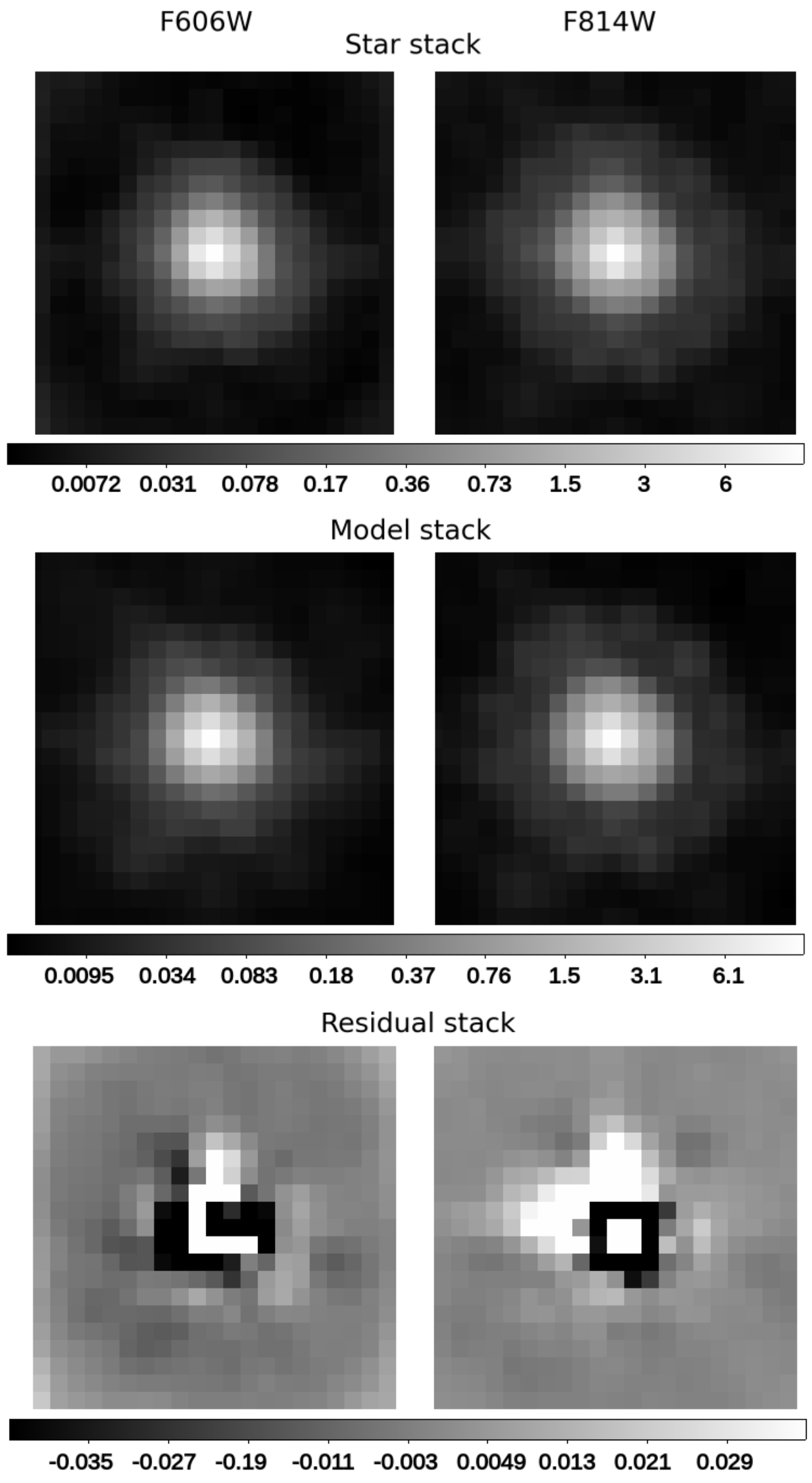

**Fig. 16.** `TinyTim` star stack, model stack, and residual stack for a star field in F606W (left panel) and in F814W (right panel) filter with a size of 21 × 21 pixels.

### 6.2. Low stellar density regime

To investigate the dependency of the precision of the inferred telescope focus on the number of available stars in a galaxy field, we create random subsamples of stars in the star fields, from which the focus is refitted. For each star field and for a range of numbers of stars within [5, 90], we employ 30 independent random subsamples. The stars selected have magnitudes in the range [22, 25], size in pixels in the range [0.8, 5.0], and a minimum value for the S/N of 50. In addition, the objects have to be separated from each other by 1 arcsec and have as minimum value for the SExtractor (Bertin & Arnouts 1996) parameter `CLASS_STAR` 0.95 (with 0 for a galaxy and 1 for a star).

Figure 17 shows the trend of the standard deviation ($\sigma_{\Delta f}$) of the difference between the recovered focus using the bootstrapped star subsample and the estimated focus using the full sample of stars, for different sizes of the bootstrap sample for F606W and F814W and for both chips 1 and 2. In this analysis, we consider star fields having at least 90 stars in order to be able to include the same number of fields[12] in each subsample. In Table 2 we report the mean focus offset as the difference between the mean focus value, obtained by considering a subsample of $N_{\rm stars}$ and the focus accounting for the full sample of stars in the star fields and the SD of this difference.

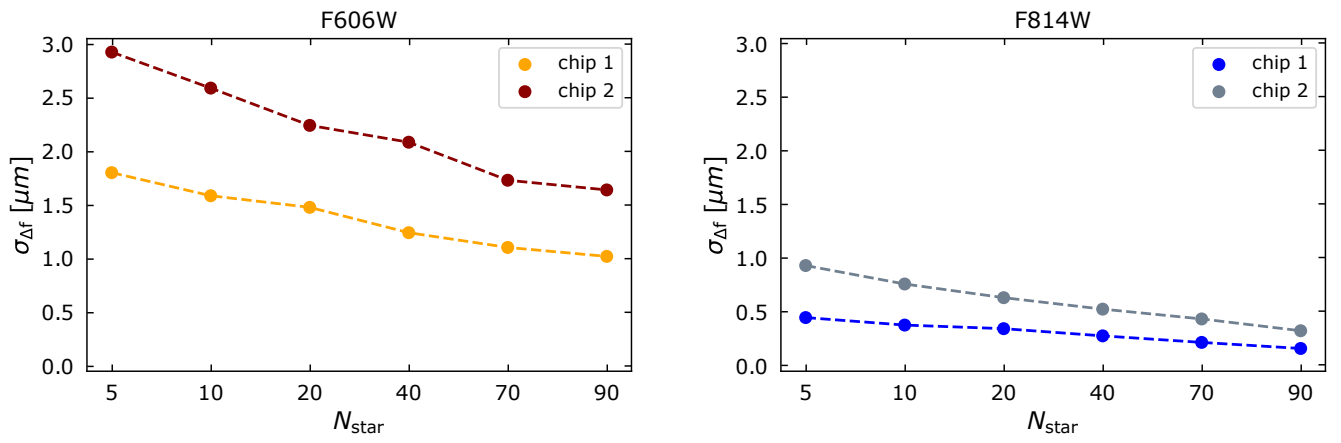

**Fig. 17.** Standard deviation ($\sigma_{\Delta f}$) of the difference between the recovered focus using bootstrapped star subsamples containing $N_{\rm star}$ stars and the focus estimated using the full sample, for different sizes of the bootstrap sample for F606W and F814W and for both chips 1 and 2.

As we expect, the greater the number of stars considered, the smaller the error in the focus estimation. The large residual in the model increases the scatter when we consider small subsamples of stars. In particular, the F606W PSF models do not fit either the core or the wings well, resulting in inaccurate focus values, with a maximum $\sigma_{\Delta f}$ of $\sim 3\ \mu$m. The problem could be due to the presence of unresolved binary stars in our sample. Indeed, this should increase the observed size of the PSF, biasing the model to a focus value which provides a larger size, meaning the best-fit focus will be pulled away from a correct value. The same behaviour is also found in Gillis et al. (2020), where it is shown that, for F606W star fields, the algorithm generally finds the best solution to the focus at larger focus offset values. Additionally, in crowded fields such as star clusters, the close proximity of stars can cause the detection of a single, but larger, star. From the visual inspection of the star fields, we note that many images are crowded star fields, especially for the F606W filter, while some observations in F814W include star fields with globular clusters. Those fields are the same which avoid the region of −8 to −2 $\mu$m in the comparison between the focus values and the principal component coefficients in Appendix B.2. We plan to investigate these hypotheses further in future analyses. We note that by combining the two chips for F606W (even if we consider $N_{\rm stars}$= 90 stars), we can recover the focus value with a scatter of $\sigma_{\Delta f} \simeq 1.4\ \mu$m only. For the F814W filter, we can obtain the focus value with less than 1 $\mu$m uncertainty with $N_{\rm stars}$= 5 stars only.

## 7. Summary and conclusions

Next-generation surveys such as *Euclid* offer a chance to revolutionise our understanding of dark matter and dark energy using

[12] The original sets of 645 star fields in F814W and 245 star fields in F606W are reduced by 33% for F814W chip 1, 11% for F814W chip 2, and 3.4% for F606W chip 1 and 6.3% for chip 2, such as the final samples we plot in Fig. 17 consists of 434, 573, 198, and 192 star fields, respectively.

**Table 2.** Mean focus offset and standard deviation (SD) across star subsamples in F606W and F814W filters.

| $N_{\rm stars}$ | $\bar{f} - f_{\rm full\ sample} \pm \sigma_{\bar{f}-f_{\rm full\ sample}}$ [$\mu$m] | | | |
|---|---|---|---|---|
| | F606W chip 1 | F606W chip 2 | F814W chip 1 | F814W chip 2 |
| 5 | 0.06 ± 1.80 | −0.15 ± 2.92 | 0.01 ± 0.45 | 0.22 ± 0.93 |
| 10 | 0.23 ± 1.58 | −0.05 ± 2.59 | −0.007 ± 0.380 | 0.11 ± 0.76 |
| 20 | 0.23 ± 1.48 | 0.02 ± 2.24 | −0.03 ± 0.35 | 0.06 ± 0.63 |
| 40 | 0.17 ± 1.24 | 0.15 ± 2.08 | −0.03 ± 0.28 | 0.01 ± 0.53 |
| 70 | 0.11 ± 1.10 | 0.17 ± 1.73 | −0.01 ± 0.22 | −0.004 ± 0.437 |
| 90 | 0.12 ± 1.02 | 0.12 ± 1.64 | −0.007 ± 0.162 | −0.01 ± 0.32 |

**Notes.** The mean focus offset is calculated as the difference between the mean focus value obtained using a subsample of $N_{\rm stars}$ (first column) and the focus value from the full sample of stars in the star fields. The SD of the difference is also provided, for the filters F606W and F814W, across both chips 1 and 2 (columns 2-5).

WL measurements thanks to the large sky coverage, large wavelength range, and high spatial resolution. Before exploiting real data, simulations are of paramount importance to investigate the impact of systematic effects and to mitigate them.

In this paper, we present the preparation of a testing environment, used to quantify the impact of uncertainties regarding the use of HST data for the generation of *Euclid* WL calibration image simulations. Using simulated data, in the simplified environment we present here, we show that the tested Euclidisation procedure can provide accurate shear calibrations if one uses either a S/N accounting for the noise correlation in the images in combination with the `lanczos15` interpolation kernel or a finer pixel scale in combination with deeper HST data. Moreover, we carried out an analysis of the accuracy of the `TinyTim` PSF models for HST/ACS, investigating also star fields with lower stellar density.

Our main findings and conclusions can be summarised as follows.

– When using the native *Euclid*/VIS and HST/ACS pixel scales in the Euclidisation set-up and no correction for the impact of noise correlations on $S/N$ estimates, the multiplicative bias difference between outputs D and E has a decreasing trend toward higher S/N , as shown in Fig. 2. In addition, we have shown in Fig. 3 that the galaxy model fits retrieve consistent estimates for the half-light radii and for the fluxes for the samples of galaxies in the D and E outputs but not for the Sérsic indices. If we use a Gaussian PSF instead of a realistic *Euclid* PSF model, the averaged multiplicative bias difference increases by 0.01 on average. Furthermore, we found that increasing the HST S/N ratio by a factor of 10, the multiplicative bias difference has a minimal dependence on S/N, as illustrated in Fig. 4.

– We attempted to compensate for the correlated noise introduced by the Euclidisation by adapting the latter so that galaxies in E and D are compared at equal correlation-aware S/N (see Fig. 9). We find that the combination of this noise correlation correction with a `lanczos15` interpolation kernel can effectively mitigate $\Delta\mu$, as demonstrated in Fig. 7. This approach lowers the observed multiplicative shear bias difference below one percent. Furthermore, we have identified that the uncertainties associated with the dispersion of the added Gaussian noise $\sigma_{\rm G}$ have a 1% impact on the observed multiplicative bias differences, as illustrated in Fig. 10 .

– When employing the default interpolation kernel (`quintic`), the use of a finer pixel scale decreases the multiplicative bias difference, converging toward zero to higher S/N, as illustrated in Fig. 11. This shows that the sampling significantly affects the Euclidisation procedure. Also in this case, the galaxy model fit provides a good agreement between the parameters of D and E galaxies, but not for the Sérsic index $n$.

– The use of an average star stack and an average model stack in the convolution and deconvolution steps within the Euclidisation set-up does not affect the shear bias measurements, at least at S/N higher than 10. This shows that further HST PSF model uncertainties will likely not affect the Euclidisation procedure significantly (see Fig. 12). In addition, we have demonstrated that the resulting additive shape measurement bias differences can be mitigated via the introduction of an additional random rotation of the galaxy images after applying the deconvolution for the ACS PSF, as shown in Fig.13.

– The accuracy of the Euclidisation set-up is not degraded if input galaxies with truncated brightness profiles are used. Weak-lensing image simulations that use actual galaxy postage stamps as input should therefore yield accurate bias calibrations that are independent of the true truncation radii of galaxy brightness profiles, as illustrated in Fig. 14.

– We extended the work from Gillis et al. (2020), who recalibrated `TinyTim` PSF models for HST/ACS in the F606W filter and tested their accuracy, to an analysis of star field observations in the F814W filter, finding a similar level of accuracy for the models. We have found that the F606W images show a broader scatter in the recovered best-fit focus compared to the images in the F814W filter, as depicted in Fig. 15. Such additional F814W PSF models will be needed in future applications of the Euclidisation set-up that aim to also emulate colour-dependent effects from the combination of F606W and F814W observations. Moreover, (as detailed in Sect. 6.2), we tested the `TinyTim` PSF model recovery in scenarios where only a few stars were available to constrain the focus. We have found that for the F606W filter even with 90 stars, the focus value has substantial statistical uncertainties ($\Delta f \simeq 0.12 \pm 1.02\,\mu$m for chip 1 and $\Delta f \simeq 0.12 \pm 1.64\,\mu$m for chip 2), while for the F814W filter, we were able to obtain a $\sigma_{\Delta f} < 1\,\mu$m using just five stars.

– In Fig. B.2, we compare `TinyTim` focus estimates to the leading coefficient in the PSF principal component analysis from Schrabback et al. (2010; 2018), finding an approximately linear relation in most of the coefficient range, along with strong deviations at extreme focus values (see Appendix B.2).

The testing environment has shown that shear biases arising from Euclidisation can be mitigated under the simplifying assumptions of this study, with isolated Sérsic galaxies. We have identified the noise of the HST images (which gets correlated by the Euclidisation), as well as the spatial sampling from HST, as key challenges to an accurate Euclidisation. Further studies, particularly those with more realistic morphologies, will be required before the Euclidisation can be used to generate calibration data satisfying the stringent requirements of *Euclid*. We anticipate that forthcoming investigations will expand upon this approach to handle more realistic galaxy shapes, potentially leveraging generative models (Scognamiglio et al., in preparation). Using the Euclidisation procedure to estimate calibration bias for the *Euclid* WL analysis requires further analysis on the impact of the correlated noise.

In the future, the same approach can be used to propagate the impact of other potential inaccuracies in the HST and data processing, such as residuals from the correction for charge transfer inefficiency (Massey 2010; Massey et al. 2014). However, we expect that their impact is small compared to the PSF model uncertainties and could therefore likely be neglected. Beyond the scope of this paper, there are other remaining issues of significant concern. These include biases resulting from shear estimation for low-resolution and/or low-S/N (< 10) galaxies, object detection, selection, colour gradient, and deblending.

*Acknowledgements.* DS acknowledges support from the European Union's Horizon 2020 research and innovation programme under grant agreement No 776247. While based at the University of Bonn, DS was a member of the International Max Planck Research School (IMPRS) for Astronomy and Astrophysics at the Universities of Bonn and Cologne. In addition, part of the research was carried out at the Jet Propulsion Laboratory, California Institute of Technology, under a contract with the National Aeronautics and Space Administration (80NM0018D0004). The Bonn group acknowledges support from the German Federal Ministry for Economic Affairs and Climate Action (BMWK) provided by DLR under projects 50QE1103, 50QE2002, and 50QE2302. TS acknowledges support provided by the Austrian Research Promotion Agency (FFG) and the Federal Ministry of the Republic of Austria for Climate Action, Environment, Energy, Mobility, Innovation and Technology (BMK) via the Austrian Space Applications Programme with grant numbers 899537, 900565, and 911971. BG thanks UK Space Agency for funding. Part of the simulations were performed on the HPC systems Raven and Cobra of the Max Planck Computing and Data Facility (MPCDF) in Garching, Germany. We also thank the anonymous referee for improving our manuscript. The Euclid Consortium acknowledges the European Space Agency and a number of agencies and institutes that have supported the development of *Euclid*, in particular the Agenzia Spaziale Italiana, the Austrian Forschungsförderungsgesellschaft funded through BMK, the Belgian Science Policy, the Canadian Euclid Consortium, the Deutsches Zentrum für Luft- und Raumfahrt, the DTU Space and the Niels Bohr Institute in Denmark, the French Centre National d'Etudes Spatiales, the Fundação para a Ciência e a Tecnologia, the Hungarian Academy of Sciences, the Ministerio de Ciencia, Innovación y Universidades, the National Aeronautics and Space Administration, the National Astronomical Observatory of Japan, the Netherlandse Onderzoekschool Voor Astronomie, the Norwegian Space Agency, the Research Council of Finland, the Romanian Space Agency, the State Secretariat for Education, Research, and Innovation (SERI) at the Swiss Space Office (SSO), and the United Kingdom Space Agency. A complete and detailed list is available on the *Euclid* web site (`www.euclid-ec.org`).

[1] Jet Propulsion Laboratory, California Institute of Technology, 4800 Oak Grove Drive, Pasadena, CA, 91109, USA
[2] Universität Bonn, Argelander-Institut für Astronomie, Auf dem Hügel 71, 53121 Bonn, Germany
[3] Universität Innsbruck, Institut für Astro- und Teilchenphysik, Technikerstr. 25/8, 6020 Innsbruck, Austria
[4] Institute for Astronomy, University of Edinburgh, Royal Observatory, Blackford Hill, Edinburgh EH9 3HJ, UK
[5] Leiden Observatory, Leiden University, Einsteinweg 55, 2333 CC Leiden, The Netherlands
[6] Mullard Space Science Laboratory, University College London, Holmbury St Mary, Dorking, Surrey RH5 6NT, UK
[7] Department of Physics, Centre for Extragalactic Astronomy, Durham University, South Road, DH1 3LE, UK
[8] Department of Physics, Institute for Computational Cosmology, Durham University, South Road, DH1 3LE, UK
[9] Departamento de Física, Faculdade de Ciências, Universidade de Lisboa, Edifício C8, Campo Grande, PT1749-016 Lisboa, Portugal
[10] Instituto de Astrofísica e Ciências do Espaço, Faculdade de Ciências, Universidade de Lisboa, Tapada da Ajuda, 1349-018 Lisboa, Portugal
[11] Université Paris-Saclay, CNRS, Institut d'astrophysique spatiale, 91405, Orsay, France
[12] ESAC/ESA, Camino Bajo del Castillo, s/n., Urb. Villafranca del Castillo, 28692 Villanueva de la Cañada, Madrid, Spain
[13] School of Mathematics and Physics, University of Surrey, Guildford, Surrey, GU2 7XH, UK
[14] INAF-Osservatorio Astronomico di Brera, Via Brera 28, 20122 Milano, Italy
[15] INAF-Osservatorio di Astrofisica e Scienza dello Spazio di Bologna, Via Piero Gobetti 93/3, 40129 Bologna, Italy
[16] IFPU, Institute for Fundamental Physics of the Universe, via Beirut 2, 34151 Trieste, Italy
[17] INAF-Osservatorio Astronomico di Trieste, Via G. B. Tiepolo 11, 34143 Trieste, Italy
[18] INFN, Sezione di Trieste, Via Valerio 2, 34127 Trieste TS, Italy
[19] SISSA, International School for Advanced Studies, Via Bonomea 265, 34136 Trieste TS, Italy
[20] Dipartimento di Fisica e Astronomia, Università di Bologna, Via Gobetti 93/2, 40129 Bologna, Italy
[21] INFN-Sezione di Bologna, Viale Berti Pichat 6/2, 40127 Bologna, Italy
[22] Max Planck Institute for Extraterrestrial Physics, Giessenbachstr. 1, 85748 Garching, Germany
[23] INAF-Osservatorio Astrofisico di Torino, Via Osservatorio 20, 10025 Pino Torinese (TO), Italy
[24] Dipartimento di Fisica, Università di Genova, Via Dodecaneso 33, 16146, Genova, Italy
[25] INFN-Sezione di Genova, Via Dodecaneso 33, 16146, Genova, Italy
[26] Department of Physics "E. Pancini", University Federico II, Via Cinthia 6, 80126, Napoli, Italy
[27] INAF-Osservatorio Astronomico di Capodimonte, Via Moiariello 16, 80131 Napoli, Italy
[28] INFN section of Naples, Via Cinthia 6, 80126, Napoli, Italy
[29] Instituto de Astrofísica e Ciências do Espaço, Universidade do Porto, CAUP, Rua das Estrelas, PT4150-762 Porto, Portugal
[30] Faculdade de Ciências da Universidade do Porto, Rua do Campo de Alegre, 4150-007 Porto, Portugal
[31] Dipartimento di Fisica, Università degli Studi di Torino, Via P. Giuria 1, 10125 Torino, Italy
[32] INFN-Sezione di Torino, Via P. Giuria 1, 10125 Torino, Italy
[33] INAF-IASF Milano, Via Alfonso Corti 12, 20133 Milano, Italy
[34] INAF-Osservatorio Astronomico di Roma, Via Frascati 33, 00078 Monteporzio Catone, Italy
[35] INFN-Sezione di Roma, Piazzale Aldo Moro, 2 - c/o Dipartimento di Fisica, Edificio G. Marconi, 00185 Roma, Italy
[36] Centro de Investigaciones Energéticas, Medioambientales y Tecnológicas (CIEMAT), Avenida Complutense 40, 28040 Madrid, Spain
[37] Port d'Informació Científica, Campus UAB, C. Albareda s/n, 08193 Bellaterra (Barcelona), Spain
[38] Institute for Theoretical Particle Physics and Cosmology (TTK), RWTH Aachen University, 52056 Aachen, Germany
[39] Institute of Space Sciences (ICE, CSIC), Campus UAB, Carrer de Can Magrans, s/n, 08193 Barcelona, Spain
[40] Institut d'Estudis Espacials de Catalunya (IEEC), Edifici RDIT, Campus UPC, 08860 Castelldefels, Barcelona, Spain
[41] Dipartimento di Fisica e Astronomia "Augusto Righi" - Alma Mater Studiorum Università di Bologna, Viale Berti Pichat 6/2, 40127 Bologna, Italy
[42] Instituto de Astrofísica de Canarias, Calle Vía Láctea s/n, 38204, San Cristóbal de La Laguna, Tenerife, Spain
[43] Jodrell Bank Centre for Astrophysics, Department of Physics and Astronomy, University of Manchester, Oxford Road, Manchester M13 9PL, UK
[44] European Space Agency/ESRIN, Largo Galileo Galilei 1, 00044 Frascati, Roma, Italy
[45] Université Claude Bernard Lyon 1, CNRS/IN2P3, IP2I Lyon, UMR 5822, Villeurbanne, F-69100, France
[46] Institute of Physics, Laboratory of Astrophysics, Ecole Polytechnique Fédérale de Lausanne (EPFL), Observatoire de Sauverny, 1290 Versoix, Switzerland
[47] UCB Lyon 1, CNRS/IN2P3, IUF, IP2I Lyon, 4 rue Enrico Fermi, 69622 Villeurbanne, France
[48] Instituto de Astrofísica e Ciências do Espaço, Faculdade de Ciências, Universidade de Lisboa, Campo Grande, 1749-016 Lisboa, Portugal
[49] Department of Astronomy, University of Geneva, ch. d'Ecogia 16, 1290 Versoix, Switzerland
[50] INAF-Istituto di Astrofisica e Planetologia Spaziali, via del Fosso del Cavaliere, 100, 00100 Roma, Italy
[51] INFN-Padova, Via Marzolo 8, 35131 Padova, Italy
[52] Université Paris-Saclay, Université Paris Cité, CEA, CNRS, AIM, 91191, Gif-sur-Yvette, France
[53] Institut de Ciencies de l'Espai (IEEC-CSIC), Campus UAB, Carrer de Can Magrans, s/n Cerdanyola del Vallés, 08193 Barcelona, Spain
[54] Istituto Nazionale di Fisica Nucleare, Sezione di Bologna, Via Irnerio 46, 40126 Bologna, Italy
[55] FRACTAL S.L.N.E., calle Tulipán 2, Portal 13 1A, 28231, Las Rozas de Madrid, Spain
[56] INAF-Osservatorio Astronomico di Padova, Via dell'Osservatorio 5, 35122 Padova, Italy
[57] Universitäts-Sternwarte München, Fakultät für Physik, Ludwig-Maximilians-Universität München, Scheinerstrasse 1, 81679 München, Germany
[58] Dipartimento di Fisica "Aldo Pontremoli", Università degli Studi di Milano, Via Celoria 16, 20133 Milano, Italy
[59] Institute of Theoretical Astrophysics, University of Oslo, P.O. Box 1029 Blindern, 0315 Oslo, Norway

[60] Felix Hormuth Engineering, Goethestr. 17, 69181 Leimen, Germany
[61] Technical University of Denmark, Elektrovej 327, 2800 Kgs. Lyngby, Denmark
[62] Cosmic Dawn Center (DAWN), Denmark
[63] Institut d'Astrophysique de Paris, UMR 7095, CNRS, and Sorbonne Université, 98 bis boulevard Arago, 75014 Paris, France
[64] Max-Planck-Institut für Astronomie, Königstuhl 17, 69117 Heidelberg, Germany
[65] Department of Physics and Astronomy, University College London, Gower Street, London WC1E 6BT, UK
[66] Department of Physics and Helsinki Institute of Physics, Gustaf Hällströmin katu 2, 00014 University of Helsinki, Finland
[67] Aix-Marseille Université, CNRS/IN2P3, CPPM, Marseille, France
[68] Université de Genève, Département de Physique Théorique and Centre for Astroparticle Physics, 24 quai Ernest-Ansermet, CH-1211 Genève 4, Switzerland
[69] Department of Physics, P.O. Box 64, 00014 University of Helsinki, Finland
[70] Helsinki Institute of Physics, Gustaf Hällströmin katu 2, University of Helsinki, Helsinki, Finland
[71] NOVA optical infrared instrumentation group at ASTRON, Oude Hoogeveensedijk 4, 7991PD, Dwingeloo, The Netherlands
[72] Centre de Calcul de l'IN2P3/CNRS, 21 avenue Pierre de Coubertin 69627 Villeurbanne Cedex, France
[73] Aix-Marseille Université, CNRS, CNES, LAM, Marseille, France
[74] Dipartimento di Fisica e Astronomia "Augusto Righi" - Alma Mater Studiorum Università di Bologna, via Piero Gobetti 93/2, 40129 Bologna, Italy
[75] Université Paris Cité, CNRS, Astroparticule et Cosmologie, 75013 Paris, France
[76] Institut d'Astrophysique de Paris, 98bis Boulevard Arago, 75014, Paris, France
[77] European Space Agency/ESTEC, Keplerlaan 1, 2201 AZ Noordwijk, The Netherlands
[78] Institut de Física d'Altes Energies (IFAE), The Barcelona Institute of Science and Technology, Campus UAB, 08193 Bellaterra (Barcelona), Spain
[79] Department of Physics and Astronomy, University of Aarhus, Ny Munkegade 120, DK-8000 Aarhus C, Denmark
[80] Space Science Data Center, Italian Space Agency, via del Politecnico snc, 00133 Roma, Italy
[81] Centre National d'Etudes Spatiales – Centre spatial de Toulouse, 18 avenue Edouard Belin, 31401 Toulouse Cedex 9, France
[82] Institute of Space Science, Str. Atomistilor, nr. 409 Măgurele, Ilfov, 077125, Romania
[83] Departamento de Astrofísica, Universidad de La Laguna, 38206, La Laguna, Tenerife, Spain
[84] Dipartimento di Fisica e Astronomia "G. Galilei", Università di Padova, Via Marzolo 8, 35131 Padova, Italy
[85] Institut für Theoretische Physik, University of Heidelberg, Philosophenweg 16, 69120 Heidelberg, Germany
[86] Institut de Recherche en Astrophysique et Planétologie (IRAP), Université de Toulouse, CNRS, UPS, CNES, 14 Av. Edouard Belin, 31400 Toulouse, France
[87] Université St Joseph; Faculty of Sciences, Beirut, Lebanon
[88] Departamento de Física, FCFM, Universidad de Chile, Blanco Encalada 2008, Santiago, Chile
[89] Satlantis, University Science Park, Sede Bld 48940, Leioa-Bilbao, Spain
[90] Centre for Electronic Imaging, Open University, Walton Hall, Milton Keynes, MK7 6AA, UK
[91] Infrared Processing and Analysis Center, California Institute of Technology, Pasadena, CA 91125, USA
[92] Universidad Politécnica de Cartagena, Departamento de Electrónica y Tecnología de Computadoras, Plaza del Hospital 1, 30202 Cartagena, Spain
[93] INFN-Bologna, Via Irnerio 46, 40126 Bologna, Italy
[94] Kapteyn Astronomical Institute, University of Groningen, PO Box 800, 9700 AV Groningen, The Netherlands
[95] Dipartimento di Fisica, Università degli studi di Genova, and INFN-Sezione di Genova, via Dodecaneso 33, 16146, Genova, Italy
[96] INAF, Istituto di Radioastronomia, Via Piero Gobetti 101, 40129 Bologna, Italy
[97] Astronomical Observatory of the Autonomous Region of the Aosta Valley (OAVdA), Loc. Lignan 39, I-11020, Nus (Aosta Valley), Italy
[98] Junia, EPA department, 41 Bd Vauban, 59800 Lille, France
[99] ICSC - Centro Nazionale di Ricerca in High Performance Computing, Big Data e Quantum Computing, Via Magnanelli 2, Bologna, Italy
[100] Instituto de Física Teórica UAM-CSIC, Campus de Cantoblanco, 28049 Madrid, Spain
[101] CERCA/ISO, Department of Physics, Case Western Reserve University, 10900 Euclid Avenue, Cleveland, OH 44106, USA
[102] Laboratoire Univers et Théorie, Observatoire de Paris, Université PSL, Université Paris Cité, CNRS, 92190 Meudon, France
[103] Dipartimento di Fisica e Scienze della Terra, Università degli Studi di Ferrara, Via Giuseppe Saragat 1, 44122 Ferrara, Italy
[104] Istituto Nazionale di Fisica Nucleare, Sezione di Ferrara, Via Giuseppe Saragat 1, 44122 Ferrara, Italy
[105] Kavli Institute for the Physics and Mathematics of the Universe (WPI), University of Tokyo, Kashiwa, Chiba 277-8583, Japan
[106] Dipartimento di Fisica - Sezione di Astronomia, Università di Trieste, Via Tiepolo 11, 34131 Trieste, Italy
[107] Minnesota Institute for Astrophysics, University of Minnesota, 116 Church St SE, Minneapolis, MN 55455, USA
[108] Université Côte d'Azur, Observatoire de la Côte d'Azur, CNRS, Laboratoire Lagrange, Bd de l'Observatoire, CS 34229, 06304 Nice cedex 4, France
[109] Institute for Astronomy, University of Hawaii, 2680 Woodlawn Drive, Honolulu, HI 96822, USA
[110] Department of Physics & Astronomy, University of California Irvine, Irvine CA 92697, USA
[111] Department of Astronomy & Physics and Institute for Computational Astrophysics, Saint Mary's University, 923 Robie Street, Halifax, Nova Scotia, B3H 3C3, Canada
[112] Departamento Física Aplicada, Universidad Politécnica de Cartagena, Campus Muralla del Mar, 30202 Cartagena, Murcia, Spain
[113] Department of Physics, Oxford University, Keble Road, Oxford OX1 3RH, UK
[114] Institute of Cosmology and Gravitation, University of Portsmouth, Portsmouth PO1 3FX, UK
[115] Department of Computer Science, Aalto University, PO Box 15400, Espoo, FI-00 076, Finland
[116] Caltech/IPAC, 1200 E. California Blvd., Pasadena, CA 91125, USA
[117] Ruhr University Bochum, Faculty of Physics and Astronomy, Astronomical Institute (AIRUB), German Centre for Cosmological Lensing (GCCL), 44780 Bochum, Germany
[118] DARK, Niels Bohr Institute, University of Copenhagen, Jagtvej 155, 2200 Copenhagen, Denmark
[119] Univ. Grenoble Alpes, CNRS, Grenoble INP, LPSC-IN2P3, 53, Avenue des Martyrs, 38000, Grenoble, France
[120] Department of Physics and Astronomy, Vesilinnantie 5, 20014 University of Turku, Finland
[121] Serco for European Space Agency (ESA), Camino bajo del Castillo, s/n, Urbanizacion Villafranca del Castillo, Villanueva de la Cañada, 28692 Madrid, Spain
[122] ARC Centre of Excellence for Dark Matter Particle Physics, Melbourne, Australia
[123] Centre for Astrophysics & Supercomputing, Swinburne University of Technology, Hawthorn, Victoria 3122, Australia
[124] School of Physics and Astronomy, Queen Mary University of London, Mile End Road, London E1 4NS, UK
[125] Department of Physics and Astronomy, University of the Western Cape, Bellville, Cape Town, 7535, South Africa

[126] ICTP South American Institute for Fundamental Research, Instituto de Física Teórica, Universidade Estadual Paulista, São Paulo, Brazil
[127] Oskar Klein Centre for Cosmoparticle Physics, Department of Physics, Stockholm University, Stockholm, SE-106 91, Sweden
[128] Astrophysics Group, Blackett Laboratory, Imperial College London, London SW7 2AZ, UK
[129] INAF-Osservatorio Astrofisico di Arcetri, Largo E. Fermi 5, 50125, Firenze, Italy
[130] Dipartimento di Fisica, Sapienza Università di Roma, Piazzale Aldo Moro 2, 00185 Roma, Italy
[131] Centro de Astrofísica da Universidade do Porto, Rua das Estrelas, 4150-762 Porto, Portugal
[132] Institute of Astronomy, University of Cambridge, Madingley Road, Cambridge CB3 0HA, UK
[133] Department of Astrophysics, University of Zurich, Winterthurerstrasse 190, 8057 Zurich, Switzerland
[134] Theoretical astrophysics, Department of Physics and Astronomy, Uppsala University, Box 515, 751 20 Uppsala, Sweden
[135] Department of Physics, Royal Holloway, University of London, TW20 0EX, UK
[136] Department of Astrophysical Sciences, Peyton Hall, Princeton University, Princeton, NJ 08544, USA
[137] Cosmic Dawn Center (DAWN)
[138] Niels Bohr Institute, University of Copenhagen, Jagtvej 128, 2200 Copenhagen, Denmark
[139] Center for Cosmology and Particle Physics, Department of Physics, New York University, New York, NY 10003, USA
[140] Center for Computational Astrophysics, Flatiron Institute, 162 5th Avenue, 10010, New York, NY, USA

## Appendix A: Signal-to-noise ratio estimation

Given an initial value for the flux of the Input galaxy pair, we tune each step of the testing environment such that the output galaxies have certain measured S/N values, according to the CCD equation (Howell 1989):

$$\mathrm{S/N} = \frac{F\ [e^-]}{\sqrt{F\ [e^-] + a^2\pi\left(\frac{R_\mathrm{e}}{l}\right)^2\left(F_\mathrm{sky}\ [e^-] + (\mathrm{ron}\ [e^-])^2\right)}}\,, \tag{A.1}$$

where $F\ [e^-]$ is the flux of the galaxy, $a$ is a multiplicative factor related to the aperture, $R_\mathrm{e}/l$ is the dimensionless half-light radius, $F_\mathrm{sky}\ [e^-]$ is the sky background level, and ron $[e^-]$ is the read-out noise. In particular, since we want to emulate direct *Euclid*-like and Euclidised images, we use parameter values that match expectations for *Euclid*, as reported in Sect. 3. We consider in our analysis the following input values of S/N $\simeq$ 10, 20, 30, and 40.

To estimate the S/N, we choose an elliptical aperture with a radius three times ($a = 3$) the mock half-light radius of the Input galaxy, $R_\mathrm{e}$. The factor of 3 was chosen as a compromise between obtaining a 'true' total magnitude and precision (Kaiser et al. 1995). Once all the steps of the procedure are taken, we calculate the output $\mathrm{S/N}_\mathrm{measured}$ using two different methods:

- Photutils Aperture. Using the Astropy `Photutils` package (Bradley et al. 2020), we consider an optimal elliptical aperture and calculate the $\mathrm{S/N}_\mathrm{measured}$ within it as the ratio `kron_flux/kron_fluxerr`. This method overestimates the lowest S/N and underestimates the highest S/N. Furthermore, the estimates of the $\mathrm{S/N}_\mathrm{measured}$ have a strong dependency on the half-light radius $R_\mathrm{e}$ and the Sérsic index $n$. In particular, the $\mathrm{S/N}_\mathrm{measured}$ for galaxies with extreme values of both parameters, e.g. $R_\mathrm{e} = 0\farcs2$ and $n = 3.0$ or $R_\mathrm{e} = 0\farcs7$ and $n = 3.0$, are far off the input $\mathrm{S/N}_\mathrm{measured}$ (see e.g. the bottom panels in Fig. A.1).
  We verify also the impact of the dilatation of the input galaxy due to the PSF after the convolutions. In this case, the radius of the aperture is $R_\mathrm{e,apert} = 3\sqrt{R_\mathrm{e}^2 + R_\mathrm{e,PSF}^2}$, where $R_\mathrm{e,PSF} = 0.085$ arcsec, corresponding to half of the Full-Width-at-Half-Maximum (FWHM) of the *Euclid* PSF (Cropper et al. 2016). We note that its maximum impact is about 4% for $R_\mathrm{e} = 0\farcs2$ and $n = 2$. The impact decreases to a minimum value of about 0.4% for $R_\mathrm{e} = 0\farcs7$ and $n = 3$, regardless of the input S/N and the output E or D. Given the minor impact we decided to not take $R_\mathrm{e,PSF}$ into account in the CCD equation.

- Elliptical aperture. As a validation, we also estimate the S/N creating a circular aperture with a radius of three times $R_\mathrm{e}$ and drawing it as an `Image` with `GalSim`. In order to transform the circular aperture into an elliptical one, we first interpolate the image using `galsim.InterpolatedImage`, second we assign the two components for the ellipticity $g_1$ and $g_2$, and then we randomly shift the centre position and apply the shear, assigning the same values we give to the Input galaxy. The $\mathrm{S/N}_\mathrm{measured}$ is calculated within this aperture, where the signal is the sum of the pixel values in the aperture, and the noise is calculated in a stripe of one thousand pixels near the bottom part of the image as the variance in those pixels.

The results are shown in Fig. A.1, illustrating in the upper panel the $\mathrm{S/N}_\mathrm{measured}$ using a customised aperture and, in the bottom panel, the results for the $\mathrm{S/N}_\mathrm{measured}$ estimate from `Photutils`, as a function of the half-light radius $R_\mathrm{e}$, for D, HST-like and E. We report the results only for the intermediate Sérsic index $n = 2$ for small ($N_\mathrm{gal} \sim 5.8 \times 10^4$) samples of galaxies with S/N $\simeq$ 30. We designed the HST-like observations to have a S/N that is approximately twice as high as the S/N for the D and E images, consistent with the recovered values. The $\mathrm{S/N}_\mathrm{measured}$ estimates computed using the elliptical apertures have a weaker dependence on $R_\mathrm{e}$ (see Fig. A.1), which is why we regard this method as our default approach for S/N computation.

## Appendix B: Considering the `TinyTim` PSF models

In this appendix, we describe additional details regarding the analysis of the `TinyTim` PSF models.

### *Appendix B.1: Quality of the fit parameters for `TinyTim` PSF models*

Gillis et al. (2020) defined and investigated different quantities regarding the accuracy of PSF model fits for WL analyses. In particular, they defined $\mathrm{X}^2$ as

$$\mathrm{X}^2 = \sum_k Z_k^2, \tag{B.1}$$

which is the sum over the set of the following eight $Z_k^2$ parameters[13]

$$Z_x^{2(-)} \simeq \sum_{i=1}^{N}\left(Q_{x,\mathrm{star},i}^{(-)} - Q_{x,\mathrm{model},i}^{(-)}\right)^4,$$
$$Z_y^{2(-)} \simeq \sum_{i=1}^{N}\left(Q_{y,\mathrm{star},i}^{(-)} - Q_{y,\mathrm{model},i}^{(-)}\right)^4,$$
$$Z_+^{2(\pm)} \simeq \sum_{i=1}^{N}\left(Q_{+,\mathrm{star},i}^{(\pm)} - Q_{+,\mathrm{model},i}^{(\pm)}\right)^2,$$
$$Z_\times^{2(\pm)} \simeq \sum_{i=1}^{N}\left(Q_{\times,\mathrm{star},i}^{(\pm)} - Q_{\times,\mathrm{model},i}^{(\pm)}\right)^2,$$
$$Z_s^{2(\pm)} \simeq \sum_{i=1}^{N}\left(Q_{s,\mathrm{star},i}^{(\pm)} - Q_{s,\mathrm{model},i}^{(\pm)}\right)^2, \tag{B.2}$$

where the sums are computed over all stars used in the analysis. The $Q$ are different quality of fit parameters, each of them is related a particular feature of the PSF and a linear combination of the normalised weighted multipole moments of the surface brightness distribution of a PSF. The relation of the $Q$ parameters to different image properties is briefly sketched in the following paragraphs. For a detailed technical description we refer to Gillis et al. (2020).

To further investigate the PSF model imperfections, we plot the dependence of the different $Q$ parameters on the best focus values in Fig. B.1. The top panels show $Q_{x,\mathrm{diff}}^{(-)}$ and $Q_{y,\mathrm{diff}}^{(-)}$, which are related to the position of the centroid. $Q_{+,\mathrm{diff}}^{(-)}$ and $Q_{+,\mathrm{diff}}^{(+)}$ are the diagonal terms of the moments matrix of the PSF, $Q_{\times,\mathrm{diff}}^{(-)}$ and $Q^{(+)}_{\times,\mathrm{diff}}$ are the other two off-diagonal terms of the matrix, $Q^{(-)}_{s,\mathrm{diff}}$ and $Q^{(+)}_{s,\mathrm{diff}}$ are related to the size estimator. The subscript 'diff' in each of these labels means that the variable expresses the difference between the value for the model and the data, averaged over all stars. The superscript refers to whether it is the (+) or (−) value, respectively, as given in Eqs. (B.2). In Fig. B.1 we show the results of images for both filters F814W and F606W and both chips 1 and 2.

[13] Here we use $Z_k^2$ as a shorthand for the set of eight parameters.

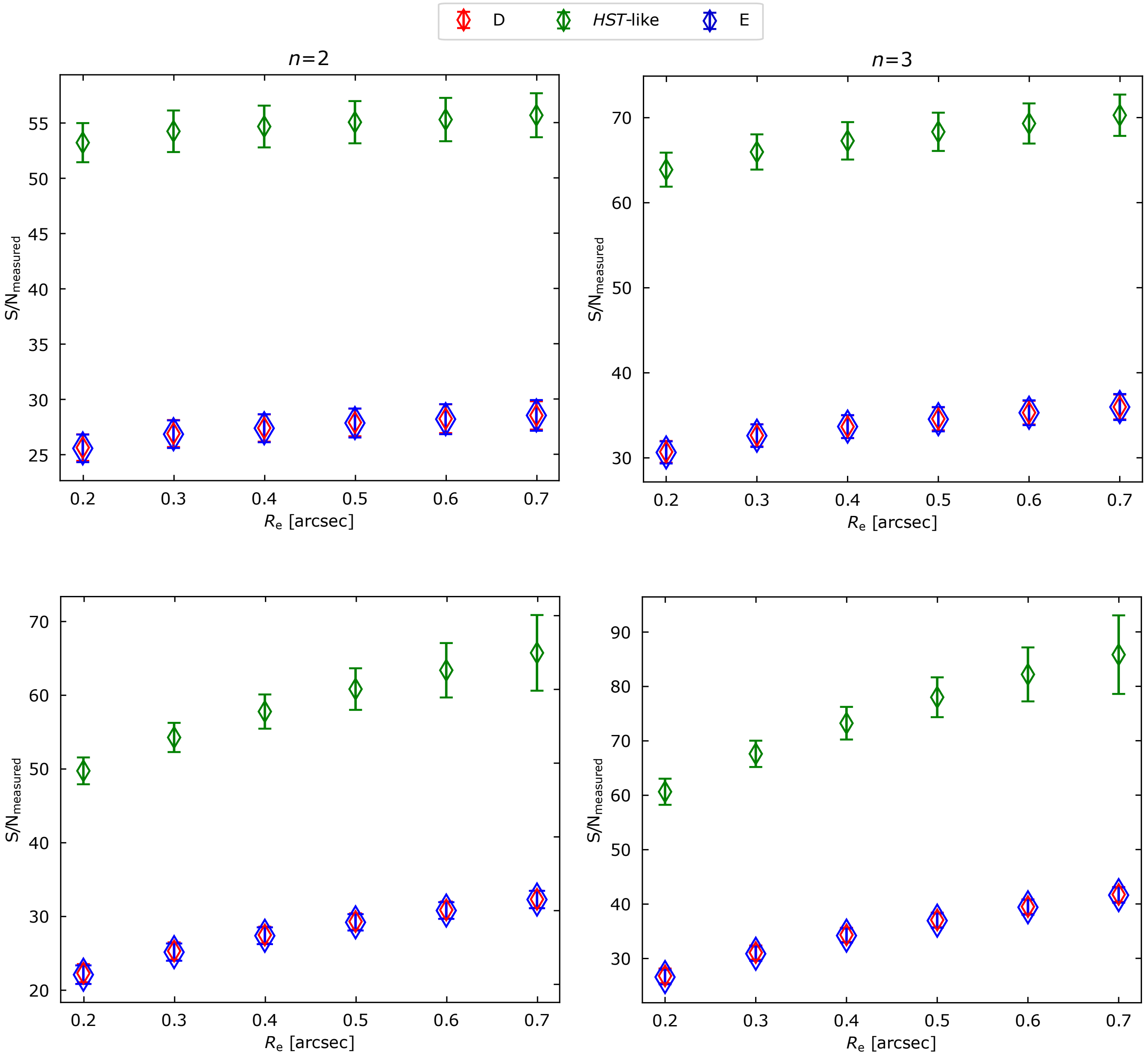

**Fig. A.1.** Comparison between the measured $S/N_{measured}$ estimated within a customised aperture (upper panels) and with `Photutils` (bottom panels) for D, HST-like, and E as a function of the half-light radius $R_e$ for an input S/N = 30. We report the results for $n = 2$ and $n = 3$. The data points are the mean and the error bars show the $1\,\sigma$ uncertainty for a number of galaxies $N_{gal} \simeq 5.8 \times 10^4$.

In the case of perfectly calibrated PSF models, we would expect that these parameters should only show some mild scatter (due to noise) around zero. Instead, some of them show significant deviations from zero, in parts focus-dependent. This was already shown by Gillis et al. (2020) for F606W, and is similarly confirmed by our analysis for F814W.

In the second- and third-row panels of Fig. B.1 we notice a gap between the values of the F606W chip 1 data versus the F606W chip 2 data. The parameters $Q_+$ and $Q_\times$ are related to the mean square contribution of the PSF shape inaccuracies to the first additive component of the shear bias. This discrepancy could be related to the temperature variations and gradients distorting the image plane in ways that are not accounted for by the `TinyTim` model. The difference between the two chips might be considered to be due to the fact that there is a vertical offset between them of about 0.5 $\mu$m (Gillis et al. 2020). This effect was also noticed by Cox & Niemi (2011), who attribute it to most likely being due to differences in spherical aberration and charge diffusion between the two chips. But we find this to likely be an insufficient explanation. Further analysis conducted by Gillis et al. (2020) shows that sometimes the model PSFs match the sizes of the observed PSF on average, but there is very large scatter in this relationship. This suggests that a possible explanation for these discrepancies might be that there is an additional spatial variation in the PSF that is not accounted for in the model.

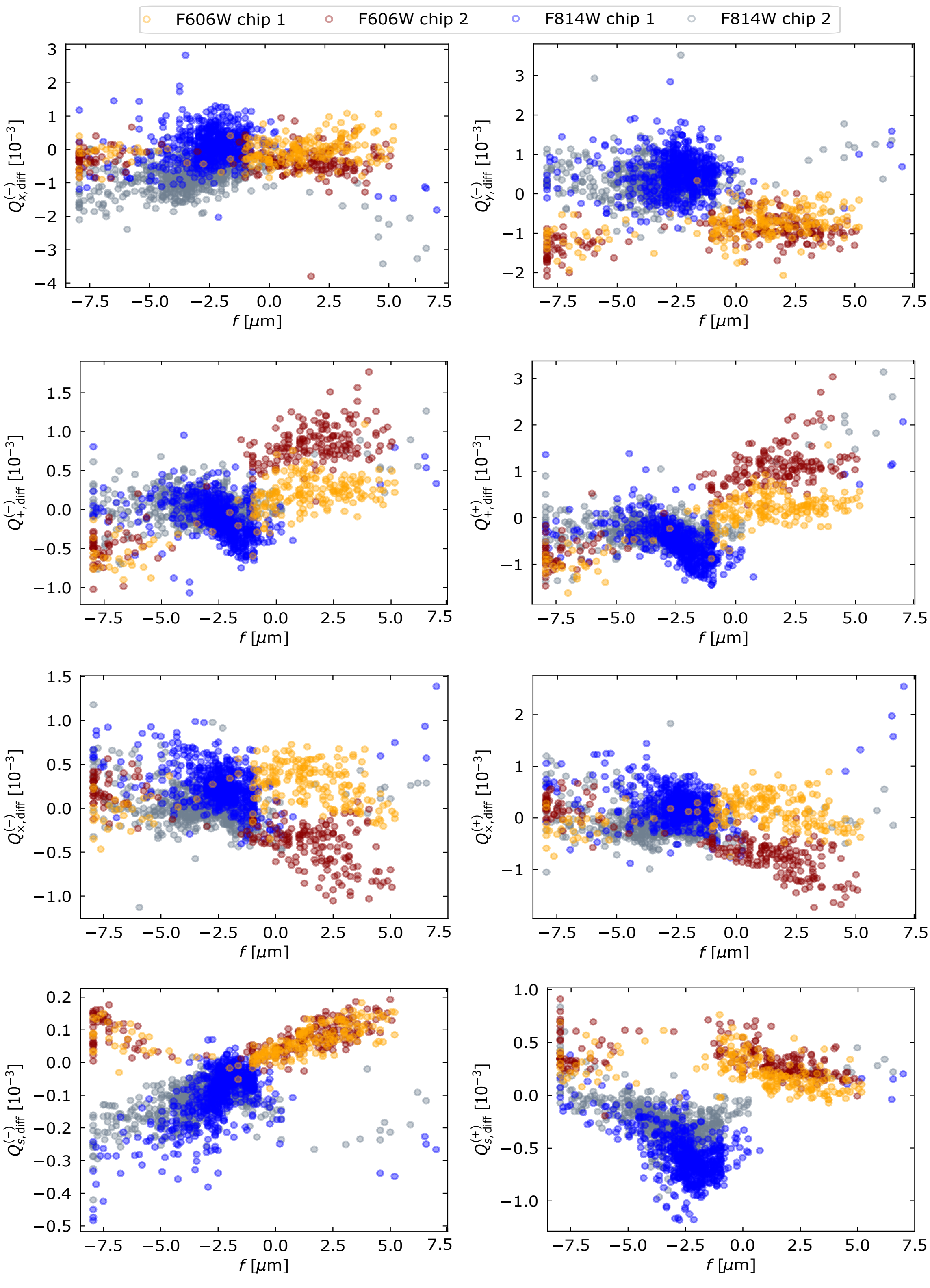

**Fig. B.1.** Best-fit focus values plotted against each component of the quality-of-fit parameter $Q$ for the filters F814W and F606W and both chips 1 and 2. In the ideal case of a perfect PSF model we should see a flat trend around zero. Some parameters exhibit a clearly different behaviour between chips 1 and 2 for the filter F606W.

### *Appendix B.2: Relation between principal component coefficients and focus values*

As an alternative approach to investigate the `TinyTim` PSF model accuracy, we use the star fields to calibrate a relation be-tween the `TinyTim` focus estimates and the first coefficient in the PCA of PSF variations from previous analyses of the same star fields conducted by Schrabback et al. (2010; 2018).

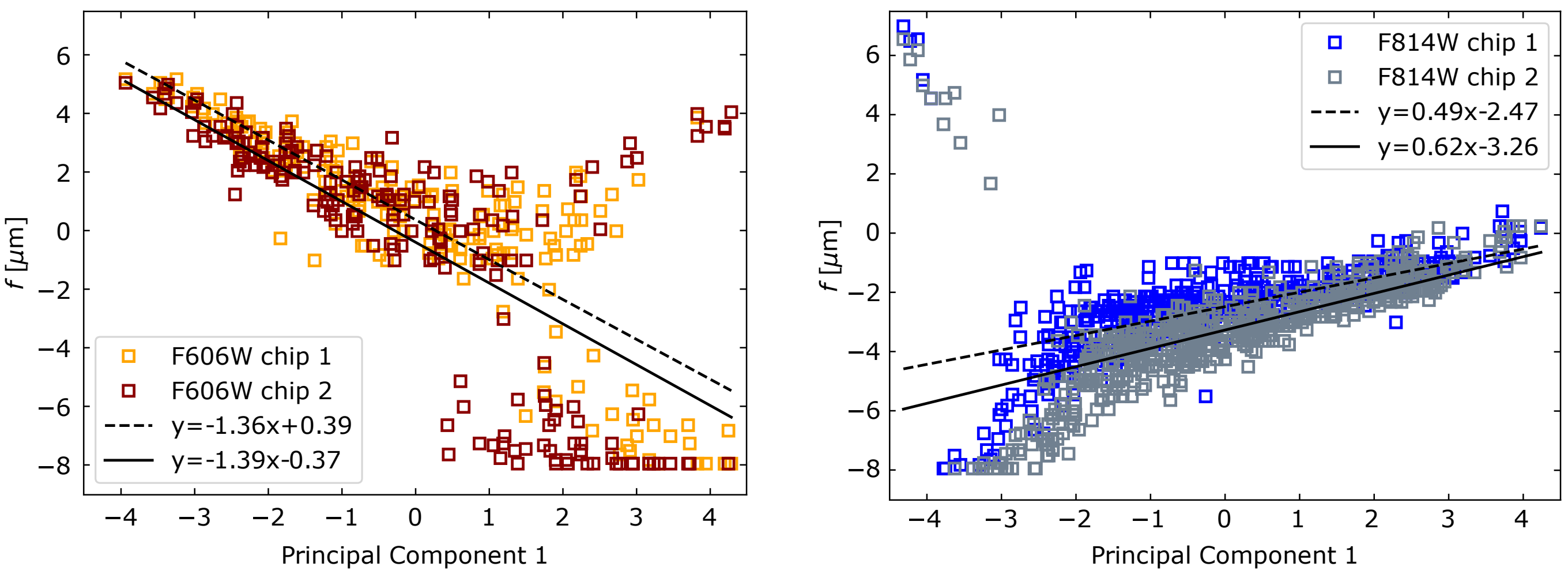

**Fig. B.2.** Best-fit focus offset values plotted against the first principal component coefficients for 205 star fields in the F606W filter (left panel) and 640 star fields in the F814W filter (right panel) and for both chips 1 and 2. The dashed and the solid lines indicate the fit lines to the data points for chip 1 and chip 2, respectively, which we also report in the legend.

Figure B.2 shows that these quantities correlate tightly in an approximately linear relation within most of the range of the first principal coefficient (PC), for both filters F606W and F814W. However, the focus values deviate strongly for high (low) PC values in F606W (F814W). For both filters, Fig. B.2 shows that some strongly negative and positive focus values are clearly off. This seems to be especially the case for fields where both chips get very different focus estimates. These large discrepancies indicate the limitation of the recalibrated `TinyTim` models in this regime, since the focus value should be similar for both chips. Depending on the required PSF model accuracy, it may therefore be necessary to drop exposures with PSF uncertainties in this regime. We also see that this problem occurs more often for F606W than for F814W.